\documentclass[a4paper,twocolumn,pra,superscriptaddress,amsmath,amssymb,showpacs,floatfix]{revtex4}
\usepackage{graphicx,epstopdf}
\usepackage{braket}

\def\openone{\leavevmode\hbox{\small1\kern-3.3pt\normalsize1}}

\def\comment#1{}

\newcommand{\be}{\begin{equation}}\newcommand{\ee}{\end{equation}}
\newcommand{\bea}{\begin{eqnarray}}\newcommand{\eea}{\end{eqnarray}}
\newcommand{\beaa}{\begin{eqnarray}}\newcommand{\eeaa}{\end{eqnarray}}
\newcommand{\ba}{\begin{array}}\newcommand{\ea}{\end{array}}
\newcommand{\bit}{\begin{itemize}}\newcommand{\eit}{\end{itemize}}
\newcommand{\ben}{\begin{enumerate}}\newcommand{\een}{\end{enumerate}}

\def\be{\begin{equation}}
\def\ee{\end{equation}}				
\def\bea{\begin{eqnarray}}				
\def\eea{\end{eqnarray}}
\def\bear{\begin{array}}				
\def\eear{\end{array}}

\def\pa{\partial}

\def\t{\tau}

\def\p{\pi}

\def\D{\Delta}
						
\def\W{\Omega}						
\def\Y{\Psi}

\def\nn{\nonumber}

\def\x5{x^{5}}

\def\fr{\frac}

\begin{document}

\title{Quantum optimal control of photoelectron spectra and angular
  distributions} 

\author{R. Esteban Goetz}
\affiliation{Theoretische Physik, Universit\"{a}t Kassel,
Heinrich-Plett-Stra{\ss}e 40, D-34132 Kassel, Germany}

\author{Antonia Karamatskou}
\affiliation{Center for Free-Electron Laser Science, DESY, Luruper
  Chaussee 149, D-22716   Hamburg, Germany}
  \affiliation{Department of Physics, Universit{\"a}t Hamburg,
    Jungiusstra{\ss}e 9, D-20355   Hamburg, Germany} 

\author{Robin Santra}
\affiliation{Center for Free-Electron Laser Science, DESY, Luruper
  Chaussee 149, D-22716   Hamburg, Germany}
  \affiliation{Department of Physics, Universit{\"a}t Hamburg,
    Jungiusstra{\ss}e 9, D-20355   Hamburg, Germany} 

\author{Christiane P. Koch}
\affiliation{Theoretische Physik, Universit\"{a}t Kassel,
Heinrich-Plett-Stra{\ss}e 40, D-34132 Kassel, Germany}
\email{christiane.koch@uni-kassel.de}

\date{\today}

\begin{abstract}
  Photoelectron spectra and photoelectron angular distributions 
  obtained in photoionization reveal important
  information on e.g. charge transfer or hole coherence in the  parent
  ion. Here we show that optimal control of the underlying quantum
  dynamics can be used 
  to enhance desired features in the photoelectron spectra and
  angular distributions. 
  To this end, we combine Krotov's method for optimal control theory with the
  time-dependent configuration interaction singles formalism and a
  splitting approach to calculate photoelectron spectra and
  angular distributions. 
  The optimization target can account for specific
  desired properties in the photoelectron angular distribution
  alone, in the photoelectron spectrum, or in both. We
  demonstrate the method for hydrogen and then apply it to 
  argon under strong XUV radiation, maximizing the
  difference of emission into the upper and lower hemispheres, in
  order to realize directed electron emission in the XUV regime. 
\end{abstract}

\pacs{32.80.Qk,32.80Rm,32.30.Jc,02.30.Yy}

\maketitle

\section{Introduction}
\label{sec:intro}

Photoelectron spectroscopy is a powerful tool 
for studying photoionization in atoms, molecules and 
solids~\cite{Hufner,MeyerPRL10,FabreAtMolPhys81,Becker,WuChemPhys11,%
BlagaNat09}. 
With the advent of new light sources, photoelectron spectroscopy using 
intense, short pulses has become available, revealing important
information about electron dynamics and  time-dependent
phenomena~\cite{KrauszModPhys09,WabnitzNAT02,CorkumNAT07,KanterPRL11}. In
particular, it allows for characterizing the light-matter interaction 
of increasingly complex systems~\cite{Hufner,WuChemPhys11,FabreAtMolPhys81}. 
Photoelectron spectra  (PES) and photoelectron angular distributions
(PAD) contain not only fingerprints  of the interaction of the
electrons with the electromagnetic fields, but also of their  
interaction and their correlations with each
other~\cite{SchwarzElecSpectros80}. PAD
in particular can be used to uncover electron
interactions and correlations~\cite{DuinkerRMP82,ReidARPC03}. 

Tailoring the pulsed electric field in its amplitude, phase or
polarization allows to control the electron dynamics, with
corresponding signatures in the photoelectron
spectrum~\cite{MeierPRL94,ShenCPL02,WollenhauptARPC05,GraefePRA05,BraunJPB14}. 
While it is natural to ask how the electron dynamics is reflected in
the experimental observables---PES and
PAD~\cite{MeierPRL94,ShenCPL02,WollenhauptARPC05,GraefePRA05,BraunJPB14},
it may also be interesting to see whether one can control or
manipulate directly these observables by tailoring the
excitation pulse. Moreover, one may be interested in certain features
such as directed electron emission without analyzing all the details of
the time evolution. This is particularly
true for complex systems where it may not be easy to trace the full
dynamics all the way to the spectrum. The question that we ask here is
how to find an external field that steers the dynamics such that the
resulting photoelectron distribution fulfills certain prescribed
properties. Importantly, the final state of the dynamics does not need
to be known. The 
desired features may be reflected in the angle-integrated PES, the
energy-integrated PAD, or both.  

To answer this question,
we employ optimal control theory (OCT), using 
Krotov's monotonically convergent method~\cite{ReichJCP12} and
adapting it to the specific 
task of realizing photoelectron distributions with prescribed
features. 
The photoelectron distributions are calculated within the time-dependent
configuration interaction singles scheme (TDCIS)~\cite{GreenmanPRA10}, 
employing the splitting method for extracting the spectral components
from the outgoing wavepacket~\cite{AntoniaPRA14,AntoniaPRA15Erratum}.
While OCT has been utilized to study the quantum control of electron
dynamics before, in the framework of TDCIS~\cite{KlamrothJCP06} as
well as the multi-configurational time-dependent Hartree-Fock
(MCTDHF) method~\cite{MundtNJP09} or time-dependent density functional
theory (TDDFT)~\cite{CastroPRL12,HellgrenPRA13},  
the  PES and PAD  have not been
tackled as control targets before.  
In fact, most previous studies did not even account for the presence
of the ionization continuum. A proper representation of
the ionization continuum becomes
unavoidable~\cite{McCurdyPRA97,MartinJPhysB99,RescignoPRA2000,bachau01,McCurdyJPhysB04},
however, when 
investigating the interaction with XUV light where a single photon  is
sufficient to ionize~\cite{Greenman}, 
and it is indispensable for the full
description of photoionization experiments.

To demonstrate the versatility of our
approach, we apply it to two different control problems: (i) We
prescribe the full three-dimensional photoelectron distribution and
search for a field that produces, 
at least approximately, a given angle-integrated PES and
energy-integrated PAD. Such a detailed
control objective is rather demanding and corresponds to a difficult
control problem. (ii) We seek to maximize the relative number of
photoelectrons emitted into the upper as opposed to the lower
hemisphere, assuming that the polarization axis of the light pulse runs through 
the poles of the two hemispheres. This implies a condition on the PAD
alone, leaving 
complete freedom to the energy dependence. The corresponding control
objective leaves considerable freedom to the optimization
algorithm and the control problem becomes much simpler. 
Maximizing the relative number of
photoelectrons emitted into the upper as opposed to the
lower hemisphere corresponds to a maximization of the PAD's asymmetry.
Asymmetric photoelectron distributions arising in strong-field ionization 
were studied previously for near-infrared few-cycle pulses where the effect was 
attributed to the carrier envelope phase~\cite{RathjePRL13,ShvetsovPRA14}.
Here, we pose the question  whether it is possible to achieve
asymmetry in the  PAD for multiphoton ionization
in the XUV regime and we seek to determine the shaped pulse that
steers the electrons  into one hemisphere.
To ensure experimental feasibility of the optimized pulses, we introduce
spectral as well as amplitude constraints. We 
test our control toolbox for hydrogen and argon atoms,
corresponding to a single channel and three active channels,
respectively. These comparatively simple examples allow for a complete
discussion of our 
optimization approach, while keeping the numerical effort at an
acceptable level. 

The remainder of this paper is organized as 
follows. Section~\ref{sec:theory} briefly reviews the methodology for
describing the electron dynamics, with Sec.~\ref{subsec:tdcis}
devoted to the TDCIS method, and Sec.~\ref{subsec:wfsm} presenting 
the wave function splitting approach. Optimal control theory for
photoelectron distributions is developed in
Sec.~\ref{sec:oct}. Specifically, we introduce the optimization
functionals to prescribe a certain  PES plus  PAD and to generate directed
photoelectron emission in Sec.~\ref{subsec:optimizationproblem}.
The corresponding optimization algorithms are presented in
Sec.~\ref{subsec:krotov}, emphasizing the combination of OCT 
with the wave function splitting method. For the additional
functionality of  restricting the spectral bandwidth of the field
in the optimization, the reader is referred to
Appendix~\ref{sec:frseqrestriction}. Our numerical results are
presented in Sec.~\ref{sec:appl1} to~\ref{sec:appl3}, demonstrating, 
for hydrogen, the prescription of the  PES and
PAD in Sec.~\ref{sec:appl1} 
and  the minimization of photoelectron emission into the lower
hemisphere in Sec.~\ref{sec:appl2}. Maximization of the 
relative number of photoelectrons emitted into the upper hemisphere is
discussed for both hydrogen and argon in Sec.~\ref{sec:appl3}.
Finally, Sec.~\ref{sec:concl}  concludes.


\section{Theory}
\label{sec:theory}
In the following, we briefly review, following
Refs.~\cite{GreenmanPRA10,AntoniaPRA14}, 
the theoretical framework for describing the electron
dynamics and the interaction with strong electric fields.

\subsection{First principles calculation of the $N$-particle wave
  function: TDCIS}\label{subsec:tdcis} 
 
Our method for calculating the outgoing electron wave packet is based on the 
time-dependent configuration interaction singles (TDCIS) scheme
\cite{GreenmanPRA10,xcid}. The time-dependent Schr\"odinger equation
of the full $N$-electron system,
\begin{eqnarray}
  i \fr{\pa}{\pa t} |\Y(t)\rangle &=& \hat{H}(t) |\Y(t)\rangle \,, \label{schr}
\end{eqnarray} 
is solved numerically  using the
Lanczos-Arnoldi propagator~\cite{leforestierJCP91,kosloffRPC94}.
To this end, the $N$-electron wave function is expanded in the
one-particle--one-hole basis: 
\begin{eqnarray}
  |\Y(t) \rangle &=&\alpha_0(t) |\Phi_0 \rangle+ \sum_{i,a}\alpha_i^a(t)|\Phi_i^a \rangle,
\label{wavefunction}
\end{eqnarray}
where the index $i$ denotes an initially occupied orbital, $a$ stands
for a virtual orbital to which the particle can be excited and
$|\Phi_0 \rangle$ symbolizes the Hartree-Fock ground state. The
full time dependent Hamiltonian has the form 
\begin{eqnarray}
  \hat{H}(t) &=& \hat{H}_0 +\hat{H}_1+\hat{\mathbf{p}}\cdot{\mathbf{A}}(t), \label{ham}
\end{eqnarray}
where $\hat{H}_0=\hat{T}+\hat{V}_{\rm nuc}+\hat{V}_{\rm MF}-E_{\rm
  HF}$ contains the kinetic energy $\hat{T}$, the nuclear potential
$\hat{V}_{\rm nuc}$, the potential at the mean-field level
$\hat{V}_{\rm MF}$ and the Hartree-Fock energy $E_{\rm HF}$.
$\hat{H}_1=\frac{1}{|r_{12}|}-\hat{V}_{\rm MF}$ describes the Coulomb
interactions beyond the mean-field level, and
$\hat{\mathbf{p}}\cdot\mathbf{A}(t)$ is the light-matter interaction within
the velocity form in the dipole approximation, assuming linear
polarization.  

The TDCIS approach is a multi-channel method, i.e., all ionization
channels that lead to a single excitation of the system are included
in the calculation. Since only states with total spin $S=0$ are
considered, only spin singlets occur and we denote the occupied orbitals by
$|\phi_{i}\rangle$. 
As introduced in Ref.~\cite{RohPRA06}, for each ionization channel all
single excitations from the occupied orbital $|\phi_{i}\rangle$ may be 
collected in one ``channel  wave function'': 
\begin{eqnarray}
  |\varphi_i(t)\rangle &=& \sum_a \alpha_i^a(t)|\phi_a \rangle\, 
  \label{channelwfct}
\end{eqnarray}
where the summation runs over all virtual orbitals,\ labeled with
$a$, which is a multi-index~\cite{GreenmanPRA10}. These channel wave functions  
allow to calculate all quantities in a
channel-resolved manner~\cite{AntoniaPRA14,AntoniaPRA15Erratum}. In the actual implementation, 
the orbitals in Eq.~\eqref{channelwfct} are expressed 
as a product of radial and angular
parts~\cite{AntoniaPRA14,GreenmanPRA10}, 
\begin{eqnarray}\label{eq:rep}
  \phi_a(\mathbf{r}) &=& \fr{u^{n_a}_{\ell_a}(r)}{r}\,Y^{\ell_a}_{m_a}(\vartheta_r,\varphi_r)\,,
\end{eqnarray}
where $Y^l_m$ denote the spherical harmonics and $u^n_l(r)$ is the radial part 
of the wave function which is represented on a pseudo-spectral spatial
grid~\cite{GreenmanPRA10}. 

\subsection{The  wave function splitting method}\label{subsec:wfsm}
The  PES and  PAD are calculated using the splitting method \cite{TongPRA06} 
which was implemented within the TDCIS scheme
\cite{AntoniaPRA14,AntoniaPRA15Erratum}. Briefly, in 
this propagation approach the wave function is split 
into an inner and an outer part using a smooth
radial splitting  function, 
\begin{eqnarray}
  \label{eq:split}
  \hat{S} &=& \left[1+e^{-(\hat{r}-r_c)/\D}\right]^{-1}\,, 
\end{eqnarray}
where the parameter $\Delta$ controls how steep the slope of the
function is and $r_c$ is the splitting radius. The channel wave
functions~\eqref{channelwfct} are used to calculate the  
spectral components in a channel-resolved manner by projecting the
outer part onto Volkov states, $| \mathbf{p}\,\!^V\rangle=
(2\p)^{-3/2}e^{i\mathbf{p}\cdot\mathbf{r}}$. To this  end, each
channel wavefunction is split into an inner and an outer part at every  
splitting time $t_j$,
\begin{subequations}
\begin{eqnarray}
  |\varphi_i(t_j)\rangle &=& |\varphi_{i,\rm in}(t_j)\rangle 
  +  |\varphi_{i,\rm out}(t_j)\rangle     
\label{eq:composite}
\end{eqnarray}
where
\begin{eqnarray}
  |\varphi_{i,\rm in}(t_j)\rangle &=& (1-\hat{S})|\varphi_i(t_j)\rangle 
\label{eq:composite2}
\end{eqnarray}
and
\begin{eqnarray}
  |\varphi_{i,\rm out}(t_j)\rangle &=& \hat{S}|\varphi_i(t_j)\rangle 
\label{eq:composite3}
\end{eqnarray}
\end{subequations}
At each splitting time, the  inner  
part, $|\varphi_{i,\rm in}(t_j)\rangle$, is 
represented in the CIS basis and further propagated with the full 
Hamiltonian~(\ref{ham}), whereas the outer part is stored and propagated 
analytically to large times with the Volkov Hamiltonian,
\begin{eqnarray}
  \hat{H}_V(\tau) &=& \frac{1}{2} \left[ \hat{\mathbf{p}}+\mathbf{A}(\t)
\right]^2\,. \label{volkovham} 
\end{eqnarray}
In this way, the outer part of the wave function can be analyzed separately in 
order to obtain information on the photoelectron. Furthermore, since the
outgoing part of the wave function is absorbed efficiently at the
splitting times, large box sizes are avoided in the inner region. 

The spectral coefficient $\varphi_i(\mathbf{p},T;t_j)$ for
a given channel $i$, originating from splitting time $t_j$ 
and evaluated at the final time $t=T$ is obtained as a function of the
momentum vector $\mathbf{p}$~\cite{AntoniaPRA14}, 
\begin{widetext}
\begin{eqnarray}
\varphi_{i,\rm out}(\mathbf{p},T;t_j) = 
\int d^3p'\langle  \mathbf{p}\,^V|\hat{ U}_V(T,t_j)| 
\mathbf{p}\,'\,\!^V\rangle \langle \mathbf{p}\,'\,\!^V | 
\varphi_{i,\rm out}(t_j)\rangle     
  =  \dfrac{2}{\pi}e^{-i\vartheta_V(\mathbf{p})}\sum_a
 (-i)^{l_a}\beta_i^a(t_j) Y^{l_a}_{m_a}(\Omega_{\mathbf{p}})     
 \int_0^{\infty} dr\, r u^{n_a}_{l_a}(r)j_{l_a}(pr)  \,,
\label{eq:Ci;tj}
\end{eqnarray}
\end{widetext}
where $\vartheta_V(\mathbf{p})$ denotes the Volkov phase,\ given by
\begin{eqnarray}
  \vartheta_V(\mathbf{p}) &=& \fr{1}{2}\int_{t_j}^T d\tau\,
  \left[\mathbf{p}+\mathbf{A}(\tau) \right]^2 \,,
\end{eqnarray}
the sum runs over the virtual orbitals, 
$\beta^a_i(t_j)$ is the overlap of the outer part with the virtual
orbital, 
\begin{eqnarray}
  \label{eq:beta}
  \beta^a_i(t_j) &=& \langle \phi_a|\varphi_{i,\rm out}(t_j)\rangle \,,
\end{eqnarray}
and $j_{l}(x)$ is the $l$th Bessel function.
$\hat{ U}_V(t_2,t_1)=\exp\left(-i \int_{t_1}^{t_2}  \hat{H}_V(\tau)
  d\t \right)$ is the evolution operator associated  
with the Volkov Hamiltonian \eqref{volkovham} and $T$ is a
sufficiently long time  
so that all parts of the wave function that are of interest have reached the 
outer region and are included in the PES. The contributions from all
splitting times must be added up coherently to form the  
total spectral coefficient for the channel $i$,
\begin{eqnarray}
  \tilde{\varphi}_{i,\rm out}(\mathbf{p},T) &=&\sum_{t_j}\varphi_{i,\rm out}(\mathbf{p},T;t_j).
\label{eq:contribution}
\end{eqnarray}
Finally, incoherent summation over all possible ionization channels
yields the total spectrum~\cite{AntoniaPRA14}, 
\begin{eqnarray}
  \fr{d^2                                                  
      \sigma  
      (\mathbf{p})}{dp\, d\W}= 
    \big| \tilde{\varphi}_{\rm out}(\mathbf{p},T) \big|^2
    = \sum_i \big| \tilde{\varphi}_{i,\rm out}(\mathbf{p},T) \big|^2\,.
\label{eq:dE}
\end{eqnarray}
The energy-integrated PAD is given by integrating over energy or, equivalently,
momentum, 
\begin{subequations}
  \label{eq:PAD_PES}
  \begin{eqnarray}\label{eq:PAD}
  \dfrac{d        
    \sigma  
  }{d\Omega} &=& 
  \int^{\infty}_0 \dfrac{d^2        
    \sigma  
    (\mathbf{p})}{dpd\Omega}\,p^2
  dp\,.
\end{eqnarray}
Analogously, the angle-integrated PES is obtained upon integration over the solid angle,
\begin{eqnarray}\label{eq:PES}
  \dfrac{d        
    \sigma  
  }{dE} &=& p \int^{2\pi}_0\int^{\pi}_0
  \dfrac{d^2        
    \sigma  
    (\mathbf{p})}{dpd\Omega}\, 
  \sin\theta d\theta d\phi
\end{eqnarray}
\end{subequations}
with $p = \sqrt{2E}$. The optimizations considered below
are based on these measurable quantities. 


\section{Optimal control theory}
\label{sec:oct}

\subsection{Optimization problem}

\label{subsec:optimizationproblem}

Our goal is to find a vector potential, or control, $\mathbf{A}(t)$, 
that steers the system from the ground state
$|\Psi(t=0)\rangle=|\Phi_0\rangle$, 
defined in  Eq.~\eqref{wavefunction}, to an unknown
final state $|\Psi(T)\rangle$ whose  PES and/or PAD
display certain desired features. Such an optimization target is
expressed mathematically as a final time functional
$J_T[\tilde{\varphi}_{\rm out},\tilde{\varphi}_{\rm
  out}^\dagger]$~\cite{ReichJCP12}. We consider 
two different final time optimization functionals in the following. 

As a first example, we seek to prescribe the angle-integrated PES and
energy-integrated PAD together. The
corresponding  final time cost functional is defined as
\begin{eqnarray}
\label{JT1} 
J^{(1)}_{T}[\tilde{\varphi}_{\rm out}(T),\tilde{\varphi}^\dagger_{\rm
  out}(T)]
&=&            
  \lambda_{1} \int \left(\tilde{\sigma}(\mathbf{p},T))
  -\tilde{\sigma}_0(\mathbf{p})\right)^2\, d^3 p \,,\quad\quad
\end{eqnarray}
where $\tilde{\sigma}(\mathbf{p},T)=d^2        
\sigma  
(\mathbf{p})/dp\, d\W$
denotes the actual photoelectron distribution, cf. Eq.~\eqref{eq:dE}, 
$\tilde{\sigma}_0(\mathbf{p})$ stands for the target distribution, and
$\lambda_{1}$ is a weight that stresses the 
importance of $J^{(1)}_{T}[\tilde{\varphi}_{\rm out},\tilde{\varphi}^\dagger_{\rm out}]$
compared to additional terms in the total optimization functional. 
The goal is thus to minimize the squared Euclidean distance between
the actual and the desired photoelectron distributions.  

Alternatively, we would like to control the difference in the number
of electrons emitted into the lower and upper hemispheres. 
This can be expressed via the following final-time functional
\begin{eqnarray}
\label{JT2}  
&&J^{(2)}_{T}[\tilde{\varphi}_{\rm out}(T),\tilde{\varphi}^\dagger_{\rm out}(T)] =\\
\nonumber
&&\quad\quad\quad\quad\lambda_2^{(-)} \int^{\pi}_{\pi/2} \sin\theta\,d\theta\int_0^{+\infty}
\big|\tilde{\varphi}_{\rm out}(\mathbf{p},T)\big|^2 p^2\, dp\\ \nonumber
&&\quad\quad\quad\quad+ \lambda_2^{(+)}\int^{\pi/2}_{0} \sin\theta\,d\theta\int_0^{+\infty} 
\big|\tilde{\varphi}_{\rm out}(\mathbf{p},T)\big|^2 p^2\, dp\\ \nonumber
&&\quad\quad\quad\quad+\lambda_2^{tot} \int^{\pi}_{0}
\sin\theta\,d\theta\int_0^{+\infty} 
\big|\tilde{\varphi}_{\rm out}(\mathbf{p},T)\big|^2 p^2\, dp\,,
\end{eqnarray}
where the first and second term correspond
to the probability of the
photoelectron being emitted into the lower and upper hemisphere,
whereas the third term is the total ionization probability. 
$\lambda_2^{(-)}$, $\lambda_2^{(+)}$ and $\lambda_2^{tot}$ are weights. The 
factor of $2\pi$ 
resulting from integration over the azimuthal angle has been absorbed into the weights. 
Directed emission can be achieved in several ways---one can 
suppress the emission of the photoelectron into the lower hemisphere,
without imposing any specific constraint on the  
number of electrons emitted into the upper 
hemisphere. This is achieved by choosing
$\lambda^{(+)}_2=\lambda^{tot}_2=0$ and  
$\lambda^{(-)}_2>0$. Alternatively, one can maximize the
difference in the number of 
electrons emitted into the upper and lower hemispheres. To this end,
the relative weights need to be chosen such that 
$\lambda^{(-)}_2>0$ and $\lambda^{(+)}_2<0$. If $\lambda^{tot}_2=0$,
the optimization seeks to increase the \textit{absolute} difference in
the number of electrons emitted into the upper and lower
hemisphere. Close to an optimum, this may result in a strong increase
in the overall ionization probability, accompanied by a very small
increase in the difference, since only the complete functional is
required to converge monotonically, and not each of its parts. 
This undesired behavior can be avoided by maximizing the relative
instead of the absolute  difference of electrons emitted into the
upper and lower hemispheres. It requires
$\lambda^{tot}_2>0$, i.e., minimization of the total
ionization probability in addition to maximizing the difference. 
Note that $\lambda_2^{tot}$ could also be absorbed
into the weights for the hemispheres, 
\begin{eqnarray}
\label{eq:JT_2_aux_by_parts_2}  
&&J^{(2)}_{T}[\tilde{\varphi}_{\rm
  out}(T),\tilde{\varphi}^\dagger_{\rm out}(T)] =
\\\nonumber                                      
&&\quad\quad\quad\quad +\lambda^{(-)}_{eff}  
\int^{\pi}_{\pi/2} \sin\theta\,d\theta\int_0^{+\infty}
\big|\tilde{\varphi}_{\rm out}(\mathbf{p},T)\big|^2 p^2\, dp
\\ \nonumber
&&\quad\quad\quad\quad+\lambda^{(+)}_{eff}  
\int^{\pi/2}_{0} \sin\theta\,d\theta\int_0^{+\infty}
\big|\tilde{\varphi}_{\rm out}(\mathbf{p},T)\big|^2 p^2\, dp
\,,
\end{eqnarray}
where $\lambda^{(+)}_{eff}
=-|\lambda_2^{(+)}|+|\lambda^{tot}_2|$ and $\lambda^{(-)}_{eff}
=|\lambda_2^{(-)}|+|\lambda^{tot}_2|$ are effective weights.
Since $\lambda^{(+)}_{eff}<0$ and $\lambda^{(-)}_{eff}>0$ in order to
maximize (minimize) emission into the upper (lower) hemisphere,  the
weights need to fulfill the condition
$|\lambda^{(+)}_2| >|\lambda^{tot}_2|$. 

The complete functional to be minimized, 
\begin{eqnarray}
  J &=& J_T[\tilde{\varphi}_{\rm out}(T),\tilde{\varphi}^\dagger_{\rm out}(T)] + C[\mathbf{A}] \,,
\label{eq:Jtotal}
\end{eqnarray}
also includes constraints $C[\mathbf{A}]$ to ensure that the control
remains finite or has a limited spectral bandwidth. The 
constraints may be written for the electric field $\mathbf{E}(t)$
associated with  the vector potential $\mathbf{A}(t)$, even 
though the minimization problem is expressed in terms of $\mathbf{A}(t)$ and the
dynamics is generated by $\hat{H}[\mathbf{A}]$, cf. Eq.~\eqref{ham}. 
The relation between the vector potential $\mathbf{A}(t)$ and the electric field
$\mathbf{E}(t)$ is given by 
\begin{eqnarray}
  \mathbf{A}(t) &=& -\int_{t_0}^t \mathbf{E}(\tau)\,d\tau\,.
\label{eq:E_A}
\end{eqnarray}
with $\mathbf{A}(t_o)=\mathbf{0}$. Without loss of generality, we can write
\begin{eqnarray}
  C[\mathbf{A}] &=& C_a[\mathbf{A}] + C_{\omega}[\mathbf{A}] + C_e[\mathbf{A}],\
\label{eq:C_def}
\end{eqnarray}
where the independent terms in the rhs. of Eq.~\eqref{eq:C_def} are
defined below. 

The first property that the optimized electric field 
must fulfill is that its integral over time vanishes, i.e., 
\begin{eqnarray}
  \int_{t_0}^{T} \mathbf{E}(t)\, dt &=& 0\,, 
  \label{eq:E_maxwell}
\end{eqnarray}
which implies, according to Eq.~\eqref{eq:E_A}, $\mathbf{A}(T)=\mathbf{A}(t_0)=0$. 
Therefore, we choose initial guess fields with $\mathbf{A}(T)=\mathbf{A}(t_0)=0$ and utilize 
\begin{eqnarray}
  C_a[\mathbf{A}] &=& \lambda_a\int s^{-1}(t)
  \left(\mathbf{A}(t)- \mathbf{A}_{\rm ref}(t)\right)^2\, dt
\label{eq:Ca}
\end{eqnarray}
with $s(T)=0$ to ensure that Eq.~\eqref{eq:E_maxwell} is fulfilled. In Eq.~\eqref{eq:Ca}, 
$\mathbf{A}_{\rm ref}(t)$ and $s(t)$ refer to a reference vector potential
and a shape function, respectively,  and $\lambda_a\ge 0$ is a weight
that stresses the importance of $C_a[\mathbf{A}]$ compared to all other
terms in the complete functional, Eq.~\eqref{eq:Jtotal}. The shape
function, $s(t)$, can be 
used to guarantee  that the control is smoothly switched on and off at
the initial and final times.

A second important property of the optimized field is a limited
spectral bandwidth. Typically, optimization without spectral
constraints leads to  
pulses with unnecessarily broad spectra which would be very hard or
impossible to produce experimentally.
To restrict the bandwidth of the electric field, $\mathbf{E}(t)$,\ we
construct a constraint $C_{\omega}[\mathbf{A}]$ in frequency domain, 
\begin{eqnarray}
  C_{\omega}[\mathbf{A}] &=& \lambda_{\omega} \int
  \tilde{\gamma}(\omega)\big|\tilde{\mathbf{E}}(\omega) \big|^2\,d\omega
  \nonumber \\
  &=& \lambda_{\omega} \int
  \tilde{\gamma}(\omega)\omega^2\big|\tilde{\mathbf{A}}(\omega)
  \big|^2\,d\omega  \,,
\label{eq:Comega}  
\end{eqnarray}
with $\tilde{\mathbf{E}}(\omega)$ being the Fourier transform of the field, 
\begin{eqnarray}
  \tilde{\mathbf{E}}(\omega) &=& \int \mathbf{E}(t)\, e^{-i\omega t}\, dt\,.
\label{eq:fourier_transform}
\end{eqnarray}
Constraints of the form of Eq.~\eqref{eq:Comega} were previously
discussed in Refs.~\cite{PalaoPRA13,ReichJMO14}: The kernel
$\tilde{\gamma}(w)$ plays a role similarly to the inverse shape
function $s^{-1}(t)$ in Eq.~\eqref{eq:Ca}, that is, it 
takes large values at all undesired frequencies. Additionally, we assume that
the symmetry requirement 
$\tilde{\gamma}(\omega) = \tilde{\gamma}(-\omega)$ is
fulfilled, see Appendix~\ref{sec:frseqrestriction} for details.

Finally, in view of experimental feasibility, we would also like to
limit the amplitude of the electric field to reasonable values. To
this end, we construct a constraint that penalizes changes in the
first time derivative of $\mathbf{A}(t)$. In fact, since
$\mathbf{E}(t)=-\dot{\mathbf{A}}(t)$, large 
values in the derivative of the vector potential translate into large
amplitudes of the corresponding electric field $\mathbf{E}(t)$. 
To avoid this, we adopt here a modified regularization
condition~\cite{HohenesterPRA07} for $\mathbf A(t)$, defining 
\begin{eqnarray}
  C_e[\mathbf{A}] &=& \nonumber
  \lambda_e\int s^{-1}(t)|{\mathbf{E}}(t)|^2\, dt\\
  &=& \lambda_e\int s^{-1}(t)|\dot{\mathbf{A}}(t)|^2\, dt \,.
  \label{eq:Ce}
\end{eqnarray}
$C_e[\mathbf{A}]$ plays the role of a penalty
functional~\cite{HohenesterPRA07}, ensuring the regularity of $\mathbf{A}(t)$,
and, as a consequence, penalizing large values on the electric field amplitude
$\mathbf{E}(t)$. The choice of the same $s^{-1}(t)$ in both Eq.~\eqref{eq:Ca} and
Eq.~\eqref{eq:Ce}  will simplify the
optimization algorithm as shown below.

\subsection{Krotov's method combined with  wave function splitting} 
\label{subsec:krotov}

Krotov's method for quantum optimal control provides a recipe to construct
monotonically convergent optimization algorithms, depending on the
target functional and additional constraints, the type of
equation of motion, and the power of the control in the light-matter
interaction~\cite{ReichJCP12}. The optimization algorithm consists of 
a set of coupled equations for the update of the control, the 
forward propagation of the state and the backward propagation of the 
so-called co-state. This set of equations needs to be solved iteratively. 
The final-time target functional (or, more
precisely, its functional derivative with respect to the propagated
state, evaluated  at the final time, which 
reflects the extremum condition on the optimization
functional~\cite{PalaoPRA03})  
determines the ``initial'' 
condition, at final time, for the backward
propagation of the co-state~\cite{ReichJCP12}. Additional constraints
which depend on the control such as those in Eq.~\eqref{eq:C_def} show
up in the update equation for the
control~\cite{ReichJCP12,ReichJMO14}. 
The challenge when combining Krotov's method with the wave function
splitting approach is due to the fact that splitting in the forward
propagation of the state implies ``glueing'' in the backward propagation
of the co-state. Here, we present an extension of the optimization
algorithm obtained with Krotov's method that takes the splitting procedure
into account. 

Evaluating the prescription 
given in Refs.~\cite{ReichJCP12,ReichJMO14}, we find for the update
equation, with $k$ labeling the iteration step, 
\begin{subequations}\label{eq:mykrotov}
\begin{eqnarray}
\label{eq:mykrotova}
  \mathbf{A}^{(k+1)}(t) &= &\mathbf{A}^{(k)}(t)
  + I^{(k+1)}(t) \\ \nonumber 
  &&- \fr{\tilde{\lambda}_{\omega}}{\lambda_a} s(t)\mathbf{A}^{(k+1)}\star h(t)
  + \fr{\lambda_e}{\lambda_a}\ddot{\mathbf{A}}^{(k+1)}(t)  \,,
\end{eqnarray}
with $\tilde{\lambda}_{\omega} =\sqrt{2\pi}\lambda_{\omega}$. 
$\mathbf{A}^{(k+1)}\star h(t)$ denotes the convolution of $\mathbf{A}^{(k+1)}$ 
and $h(t)$,
\begin{eqnarray}
\label{convoldef}
\mathbf{A}^{(k+1)}\star h(t) &=& \int \mathbf{A}^{(k+1)}(\tau)\, h(t-\tau)\,d\tau\,
\end{eqnarray}
with $h(t)$ the inverse Fourier transform of
$\tilde{h}(\omega)=\omega^2\tilde{\gamma}(\omega)$.
The second term in Eq.~\eqref{eq:mykrotova} is given by 
\begin{eqnarray}
  \label{eq:overlap}
  I^{(k+1)}(t) &=& \fr{s(t)}{\lambda_a} \mathfrak{Im}\left\{ \left\langle
    \chi^{(k)}(t)\left|\fr{\pa\hat{H}}{\pa\mathbf{A}} 
  \right|\Psi^{(k+1)}(t)\right\rangle\right\} \nonumber \\
  &=& \fr{s(t)}{\lambda_a}\mathfrak{Im}
  \left\{\langle \chi^{(k)}(t)|\hat{\mathbf{p}}|\Psi^{(k+1)}(t)
\rangle\right\}  \,,
\end{eqnarray}
\end{subequations}
where $|\Psi^{(k+1)}(t)\rangle$ and $|\chi^{(k)}(t)\rangle$ denote the
forward propagated state and 
backward propagated co-state at iterations $k+1$ and $k$, respectively.
The derivation of Eqs.~\eqref{eq:mykrotov} is detailed in
Appendix~\ref{sec:frseqrestriction}.
In order to evaluate Eqs.~\eqref{eq:mykrotov},
the co-state obtained at the previous iteration, $|\chi^{(k)}(t)\rangle$,
using the old control, $\mathbf A^{(k)}(t)$, must be known.
Its equation of motion is found to be~\cite{ReichJCP12}
\begin{subequations}\label{adjointsystem}
  \begin{eqnarray}
  i \fr{\pa}{\pa t} |\chi(t)\rangle &=& \hat{H}(t) |\chi(t)\rangle \,.
\label{eq:adjointequation}
\end{eqnarray} 
Just as $|Psi(t)\rangle$ is decomposed into channels wavefunctions, cf.
Eq.~\eqref{channelwfct}, so is the co-state. The  ``initial'' condition at the 
final time $T$ is written separately for each channel, 
\begin{eqnarray}
\label{initcond}
|\tilde{\chi}_{i,\rm out}(T)\rangle &=& - \fr{\pa
J_T[\tilde{\varphi}_{i,\rm out}(T),\tilde{\varphi}^\dagger_{i,\rm out}(T)]}{\pa\langle
\tilde{\varphi}_{i,\rm out}(T)|}\,.
\end{eqnarray}
\end{subequations}
Evaluation of Eq.~\eqref{initcond} requires knowledge of the outer
part of each channel wavefunction, $|\tilde{\varphi}_{i,\rm
  out}(T)\rangle$, which is obtained by forward propagation of the
initial state, including the splitting procedure.
In what follows,
$\hat{U}(t^\prime,\tau;\mathbf{A}(t))$ denotes  the evolution operator that 
propagates a given state from time $t=\tau$  to $t=t^\prime$ under the control 
$\mathbf{A}(t)$.\ We distinguish the time evolution operators for the
inner part, $\hat{U}_{F}(t^\prime,\tau;\mathbf{A}(t))$, generated by the
full Hamiltonian, Eq.~\eqref{ham},  and for the outer part, 
$\hat{U}_V(t^\prime,\tau;\mathbf{A}(t))$, 
generated by the Volkov Hamiltonian, Eq.~\eqref{volkovham}. 
For every channel, the total wavefunction is given by
\begin{eqnarray}
  |\varphi_i^{(k+1)}(t)\rangle &=&
  |\varphi_{i,\rm in}^{(k+1)}(t)\rangle + 
  |\tilde{\varphi}_{i,\rm out}^{(k+1)}(t)\rangle\,,
\label{eq:tilde1}
\end{eqnarray}
which is valid for arbitrary times
$t\ge t_1$ with $t_1$ the first splitting time.
The second term in Eq.~\eqref{eq:tilde1} reads  
\begin{eqnarray}
|\tilde{\varphi}^{(k+1)}_{i,\rm out}(t)\rangle &=&
\sum^{\lfloor t/t_1 \rfloor}_{j=1}|\varphi_{i,\rm out}^{(k+1)}(t;t_j)\rangle
\nonumber \\
&=&\sum^{\lfloor t/t_1
  \rfloor}_{j=1}\hat{U}_V(t,t_{j};\mathbf{A}^{(k+1)})\, 
|\varphi_{i,\rm out}^{(k+1)}(t_{j})\rangle\quad
\label{eq:tilde2}
\end{eqnarray}
with $\lfloor x \rfloor = \rm max \{m\in{Z},m\le x \}$. 
Equation~\eqref{eq:tilde2} accounts for  the 
fact that for $t\ge t_2$, all outer parts 
$|\varphi^{(k+1)}_{i,\rm out}(t;t_j)\rangle$ that originate at splitting times
$t_j\le t$ must be summed up coherently.
Propagation of all $|\varphi^{(k+1)}_{i,\rm out}(t;t_j)\rangle$ 
and continued splitting of $|\varphi^{(k+1)}_{i,\rm in}(t)\rangle$
eventually yields the
state at final time, $|\varphi^{(k+1)}_i(T)\rangle$. Its outer part is given by 
\begin{eqnarray}
  |\tilde{\varphi}^{(k+1)}_{i,\rm out}(T)\rangle &=& 
  \sum^{N}_{j=1}\,|\varphi^{(k+1)}_{i,\rm out}(T;t_j)\rangle\,,
\label{phiT} 
\end{eqnarray} 
where $N$ denotes the number of splitting times utilized 
during propagation, and the last splitting time $t_N$ is chosen
such that $t_N\le T$. The best compromise between size of the spatial
grid,\ time step and duration between two consecutive splitting times is
discussed in Ref.~\cite{AntoniaPRA14}.

Equation~\eqref{initcond} can now be evaluated: Since our final time
functionals all involve the product  
$\tilde{\varphi}_{\rm out}(\mathbf{p},T)\cdot
\tilde{\varphi}_{\rm out}^*(\mathbf{p},T)=\sigma(\mathbf{p},T)$,\ 
Eq.~\eqref{initcond} can be written, at the $k$th iteration of the
optimization, as 
\begin{subequations}\label{eq:chiT}
  \begin{equation}                             
    \tilde{\chi}^{(k)}_{i,\rm out}(\mathbf{p},T)= \mu(\mathbf{p})\,
    \tilde{\varphi}^{(k)}_{i,\rm out}(\mathbf{p},T)    \,,
    \label{mup}
  \end{equation}
  where $\mu(\mathbf{p})$ is a function that depends on the target
  functional under consideration. 
  It becomes
  \begin{equation}
    \label{eq:mu1}
    \mu_1^{(k)}(\mathbf p)= -2\lambda_1\left( \tilde{\sigma}^{(k)}(\mathbf{p},T) -\tilde{\sigma}_0(\mathbf{p})\right)
  \end{equation}
  for $J_T^{(1)}$ given in Eq.~\eqref{JT1} and 
  \begin{equation}
    \label{eq:mu2}
    \mu_2(\mathbf p)=
    \lambda_2^{-}\, \openone_{\vartheta_-}(\theta) +
    \lambda_2^{+}\, \openone_{\vartheta_+}(\theta) 
  \end{equation}
  for $J_T^{(2)}$ given in Eq.~\eqref{JT2}.\
\end{subequations}
The intervals $\vartheta_{-}=[\pi/2,\pi]$ and
$\vartheta_{+}=[0,\pi/2]$ denote the 
lower and upper hemispheres, respectively, 
and $\openone_{\vartheta_{\pm}}(\theta)$ is the characteristic  
function on a given interval,                            
\[
  \openone_{\vartheta_{\pm}}(\theta) = 
\begin{cases}
1 & \text{if}\ \theta\in\vartheta_{\pm} \\
0 & \text{if}\ \theta\notin\vartheta_{\pm} 
\end{cases}
\]
with $\theta\in[0,\pi]$ the polar angle with respect to the polarization axis. 
According to Eqs.~\eqref{schr} and~\eqref{eq:adjointequation}, or,
more precisely, since we do not consider intermediate-time constraints
that depend on the state of the system~\cite{ReichJCP12}, 
$|\Psi(t)\rangle$ and its co-state $|\chi(t)\rangle$ obey the
same equation of motion. For that reason, it is convenient to define inner and outer parts of
$|\chi(t)\rangle$, analogously to the forward propagated state,
\begin{subequations}
  \label{eq:sub_chis}
  \begin{eqnarray}
  |\chi_i^{(k)}(t)\rangle &=&
|\chi_{i,\rm in}^{(k)}(t)\rangle + |\tilde{\chi}_{i,\rm out}^{(k)}(t)\rangle\,.
\label{eq:chi_total}
\end{eqnarray}
with
\begin{equation}
|\tilde{\chi}^{(k)}_{i,\rm out}(T)\rangle
= \sum^{N}_{j=1}\,|\chi^{(k)}_{i,\rm out}(T;t_j)\rangle\,.
\label{chiT}
\end{equation}
\end{subequations}

Eq.~\eqref{chiT} implies that also
$|\tilde{\chi}^{(k)}_{i,\rm out}(T)\rangle$ is obtained by coherently
summing up the contributions from all splitting times.

Conversely, the outer part of the co-state originating at the
splitting time $t_j$ and 
evaluated at the same time is given by  
\begin{equation}
\chi^{(k)}_{i,\rm out}(\mathbf{p},t_j;t_j) = 
\mu(\mathbf{p})\varphi^{(k)}_{i,\rm out}(\mathbf{p},t_j;t_j)\,.
\label{chitj}
\end{equation}
The next step is to construct the total co-state at an
arbitrary time $t$, $|\chi^{(k)}_i(t)\rangle$, required in
Eq.~\eqref{eq:mykrotov}, from all 
$|\chi^{(k)}_{i,\rm  out}(t_j;t_j)\rangle$ 
using Eq.~\eqref{chitj}. This is achieved by backward propagation and ``glueing''
inner and outer parts, as opposite to ``splitting'' during the forward propagation.
However, when reconstructing the co-state by backward propagation,
care should be taken to not to perform the ``glue'' procedure twice or more, at a given
splitting time.
The backward propagation of the co-state is explicitly explained in what follows: Since at the final time $T$, 
the total co-state
is given by a coherent superposition of all outer parts 
originating at the splitting times $t_j$, cf. Eq.~\eqref{chiT}, 
it suffices to store all $|\varphi^{(k)}_{i,\rm out}(t_j;t_j)\rangle$
and apply Eq.~\eqref{chitj} to   
evaluate $|{\chi}^{(k)}_{i,\rm out}(t_j;t_j)\rangle$. We recall that
$|{\chi}^{(k)}_{i,\rm out}(t_j;t_j)\rangle$, respectively $|{\phi}^{(k)}_{i,\rm out}(t_j;t_j)\rangle$,
denote the outer part born exclusively at $t=t_j$ and evaluated at the same
splitting time.
Once all outer parts of the co-state are evaluated at every splitting 
time using Eq.~\eqref{chitj}, $|\chi^{(k)}_i(t)\rangle$ is obtained for all times $t$ by 
backward propagation and ``'glueing", with the additional care of not ``glueing'' twice or more. In detail, 
$|{\chi}^{(k)}_{i,\rm out}(t_N;t_N)\rangle$ is propagated
backwards from $t_N$ to $t_{N-1}$ using the full CIS Hamiltonian,
$\hat{H}$, cf. Eq.~\eqref{ham}. 
The resulting wave function at  
$t=t_{N-1}$ is $|\chi^{(k)}_{i,\rm in}(t_{N-1})\rangle$. The outer part born
exclusively at the splitting time $t=t_{N-1}$ is obtained using  
Eq.~\eqref{chitj}, and the 
``composite'' wave function $|{\chi}^{(k)}_i(t_{N-1})\rangle$  
is obtained  by ``glueing'' $|\chi^{(k)}_{i,\rm in}(t_{N-1})\rangle$ 
and $|\chi^{(k)}_{i,\rm out}(t_{N-1};t_{N-1})\rangle$,
\[ 
  |{\chi}^{(k)}_i(t_{N-1})\rangle = |\chi^{(k)}_{i,\rm in}(t_{N-1})\rangle +
|\chi^{(k)}_{i,\rm out}(t_{N-1};t_{N-1})\rangle\,. 
\]
The procedure is now repeated: the composite co-state 
$|{\chi}^{(k)}_i(t_{N-1})\rangle$ is                         
propagated backwards from $t=t_{N-1}$ to  $t=t_{N-2}$ using the full
CIS Hamiltonian, resulting in  $|\chi^{(k)}_{i,\rm in}(t_{N-2})\rangle$, and 
``glueing'' yields the composite wave function at $t=t_{N-2}$, 
\[                                            
  |{\chi}^{(k)}_i(t_{N-2})\rangle = |\chi^{(k)}_{i,\rm in}(t_{N-2})\rangle +
|\chi^{(k)}_{i,\rm out}(t_{N-2};t_{N-2})\rangle\, ,
\]
with $|\chi^{(k)}_{i,\rm out}(t_{N-2},t_{N-2})\rangle$ given by Eq.~\eqref{chitj}; 
and so on and so forth for all splitting times $t_j$,
until $t=t_0$, where $t_0$ refers to the initial time. During the backward
propagation, as described above, the resulting co-state is stored in CIS basis. 
It gives by construction, at an arbitrary time $t$, the first term in Eq.~\eqref{eq:chi_total}.
The second term in Eq.~\eqref{eq:chi_total} involving the outer parts ``born''
at the splitting times $t=t_j$ and evaluated at $t>t_j$ is merely given by
forward propagating analytically all $|\chi_{i,\rm out}(t_j;t_j)\rangle$ using the
Volkov Hamiltonian, and summing them up coherently according to
Eq.~\eqref{eq:tilde2}. This allows to calculate the ``total'' co-state
wavefunction at an arbitrary time $t$, analogously to
$|\varphi_i(t)\rangle$. Finally, Eqs.~\eqref{eq:chi_total} and~\eqref{eq:tilde1} 
allow for evaluating Krotov's update equation for 
the control, Eq.~\eqref{eq:mykrotov}, where the iteration label just indicates
whether the guess, $\mathbf{A}^{(0)}(t)$, the old,  $\mathbf{A}^{(k)}(t)$,
or the new control, $\mathbf{A}^{(k+1)}(t)$, enter the propagation of 
$|\chi_i(t)\rangle$ and $|\varphi_i(t)\rangle$, respectively. 
A difficulty in solving the update equation for the control, 
is given by the fact that Eq.~\eqref{eq:mykrotov} is
implicit in $\mathbf{A}^{(k+1)}(t)$. Strategies to overcome this
obstacle depend on the additional constraints.

\subsection{Additional constraints}
\label{subsec:constraints}

Implicitness of Eq.~\eqref{eq:mykrotov} in $\mathbf{A}^{(k+1)}(t)$
for $\lambda_\omega=\lambda_e=0$ can easily 
be circumvented by a zeroth-order solution, employing two shifted time
grids, one for the states,\ which are evaluated at $n\Delta t$,\  and
another one for the control, which is evaluated at $(n+1/2)\Delta t$~\cite{PalaoPRA03}. 
However, for $\lambda_\omega\neq 0$,\ Eq.~\eqref{eq:mykrotov} corresponds
to a second order Fredholm equation with  inhomogeneity
$I^{(k+1)}(t)$~\cite{ReichJMO14}. Numerical solution is possible
using, for example, the method of degenerate
kernels~\cite{ReichJMO14}. To this end, the inhomogeneity $I^{(k+1)}(t)$,
which depends on $|\varphi^{(k+1)}(t)\rangle$ and thus on
$\mathbf{A}^{(k+1)}(t)$, is first approximated to  zeroth order by solving 
Eq.~\eqref{eq:mykrotov} with $\lambda_{\omega}=0$\,\ that is,\ without 
frequency constraints; and the resulting approximation 
$I_0^{(k+1)}(t)$ is then used to solve the Fredholm equation. While an
iterative procedure to improve the approximation of $I^{(k+1)}(t)$ is
conceivable, the zeroth order approximation was found to be sufficient
in Refs.~\cite{PalaoPRA13,ReichJMO14}. 
Here, we adopt a slightly different procedure,\ in the sense that the 
Fredholm equation is not solved in  time domain but in frequency
domain. This allows us to treat the cases 
$\lambda_\omega\neq 0$ and $\lambda_e\neq 0$ on the same footing. 
It is made possible by assuming that $s(t)$ in Eqs.~\eqref{eq:Ca}
and~\eqref{eq:Ce} 
rises and falls off very quickly at the beginning and end of 
the optimization time interval. This 
judicious choice of $s(t)$ together with the fact that the Fourier
transform of a convolution of two functions  in time domain, as
encountered in Eq.~\eqref{eq:mykrotov}, is the product of the functions
in frequency domain, allows to approximate 
\begin{eqnarray}
\left|
\int s(t) {\Gamma}^{(k+1)}(t)\, e^{-i\omega t} dt - S_0\int
{\Gamma}^{(k+1)}(t)\, e^{-i\omega t} dt \right| &\le &\epsilon\,,\nn \\
\label{epsilon}
\end{eqnarray}
where $\epsilon$ is a small, positive number and $\Gamma^{(k+1)}(t)$
is defined as 
\begin{eqnarray}
{\Gamma}^{(k+1)}(t) = \mathbf{A}^{(k+1)}\star h(t) \,.
\end{eqnarray}
A possible choice for $s(t)$ to fulfill the condition~\eqref{epsilon}
is  
\begin{eqnarray}
\label{choice_st}
  s(t) &=& e^{-\beta ((t-t_c)/2\sigma)^{2n}}\,,
\end{eqnarray}
where $\sigma$ refers to the duration of the pulse centered at
$t=t_c$. 
If Eq.~\eqref{epsilon} is satisfied,\ we can easily take the Fourier
transform of both sides of Eq.~\eqref{eq:mykrotova} to get
\begin{subequations}
\label{eq:newmethod}
\begin{eqnarray}
  \label{eq:fourier}
  \tilde{\mathbf{A}}^{(k+1)}(\omega) &=& 
  \dfrac{\tilde{\mathbf{A}}^{(k)}(\omega) + \tilde{I}^{(k+1)}(\omega)}
  {1+\dfrac{\tilde{\lambda}_{\omega}}{\lambda_a}\omega^2\tilde{\gamma}(\omega)
    +\dfrac{\lambda_e}{\lambda_a}\omega^2} 
\end{eqnarray}
with $\mathbf{A}^{(k+1)}(t) =\int
\tilde{\mathbf{A}}^{(k+1)}(\omega)\,e^{+i\omega t}\,d\omega/\sqrt{2\pi}$.
Note that Eq.~\eqref{eq:fourier} becomes exact if $s(t)$ is constant. 
Approximating $\tilde{I}^{(k+1)}(\omega)$ by its zeroth order solution
analogously to Ref.~\cite{ReichJMO14},  Eq.~\eqref{eq:fourier} can be
expressed as 
\begin{eqnarray}
\label{transfer}
\tilde{\mathbf{A}}^{(k+1)}(\omega) &=& 
  \tilde{G}(\omega)\,{\tilde{\mathbf{A}}}_0^{(k+1)}(\omega) \,,
\end{eqnarray}
where $\tilde{\mathbf{A}}_0^{(k+1)}(\omega)$ is the zeroth order solution of the 
updated control,\ found by solving Eq.~\eqref{eq:mykrotov} with 
$\lambda_{\omega}=\lambda_e=0$,
\begin{eqnarray}
  \tilde{\mathbf{A}}_0^{(k+1)}(\omega)&=&\tilde{\mathbf{A}}^{(k)}(\omega) 
+ \tilde{I}^{(k+1)}_0(\omega)\,,
\end{eqnarray}
and  $\tilde{G}(\omega)$ is a transfer function  given by
\begin{eqnarray}
  \tilde{G}(\omega) = \left(1+
     \dfrac{\lambda_{\omega}}{\lambda_a}\omega^2\tilde{\gamma}(\omega)
    +\dfrac{\lambda_e}{\lambda_a}\omega^2\right)^{-1}\,. 
\label{eq:Gw}
\end{eqnarray}
\end{subequations}

\subsection{Summary of the algorithm}
\label{subsec:algo}

The complete implementation of the optimization within the
time-splitting framework of the TDCIS method is summarized as follows: 
\begin{enumerate}
\item Choose an initial guess for the vector potential, $\mathbf A^{(k=0)}(t)$. 
\item Forward propagation of the state:
\begin{enumerate} 
  \item Propagate $|\Psi^{(k=0)}(t=0)\rangle$, cf. Eq.~\eqref{wavefunction}, 
  from $t=0$ until the first splitting time,
  $t=t_1$, in the CIS basis. We label the
  projection of the propagated state onto the channel wavefunctions defined in
  Eq.~\eqref{channelwfct} by $i=1,2,\dots$, 
  while $i=0$ is reserved for the projection onto the
  Hartree-Fock ground state. 
 \item At $t=t_1$, apply the splitting function defined in 
  Eq.~\eqref{eq:split} to obtain $|\varphi^{(k)}_{i,\rm in}(t_1)\rangle$ and 
  $|\varphi^{(k)}_{i,\rm out}(t_1;t_1)\rangle$. Store the outer part in the CIS
  representation, before transforming
  $|\varphi^{(k)}_{i,\rm out}(t_1;t_1)\rangle$ to the Volkov representation.  
\item Propagate $|\varphi^{(k)}_{i,\rm in}(t_1)\rangle$ using $\hat{H}$ and 
 $|\varphi^{(k)}_{i,\rm out}(t_1;t_1)\rangle$ using $\hat{H}_V$ from
 $t=t_1$ to the next splitting time, $t=t_2$.
\item\label{item:split} At $t=t_2$, apply the splitting function to
  $|\varphi^{(k)}_{i,\rm in}(t_2)\rangle$, again store 
  the resulting outer part in CIS representation,\ and
  transform $|\varphi^{(k)}_{i,\rm out}(t_2;t_2)\rangle$ to the Volkov
  representation. 
\item\label{item:prop} Propagate $|\varphi^{(k)}_{i,\rm in}(t_2)\rangle$ using 
  $\hat{H}$ and  $|\tilde{\varphi}^{(k)}_{i,\rm out}(t_2)\rangle= 
  |\varphi^{(k)}_{i,\rm out}(t_2;t_1)\rangle+ |\varphi^{(k)}_{i,\rm out}(t_2;t_2)\rangle $ 
  using $\hat{H}_V$ from  $t=t_2$ to the next splitting time $t=t_3$.
\item Repeat steps~\eqref{item:split} and~\eqref{item:prop} for all
  remaining splitting times $t_j$ up to $t_N$.
\item Propagate  for each channel wave function,\ $|\tilde{\varphi}^{(k)}_{i,\rm out}(t_N)\rangle = 
  \sum_{t_j=t_1}^{t_N}|\varphi^{(k)}_{i,\rm out}(t_N;t_j)\rangle$ from the
  last splitting time, $t=t_N$, to the final time, $t=T$, to obtain
  $|\tilde{\varphi}^{(k)}_{i,\rm out}(T)\rangle$ and evaluate the target functional $J_T$.
\item Calculate $\chi^{(k)}_{i,\rm out}(\mathbf{p},T)$ according to
  Eq.~\eqref{mup}. 
\end{enumerate}
\item\label{item:backw} Backward propagation of the co-state:
\begin{enumerate}
\item Calculate  $\mu(\mathbf{p},T)$  according to Eq.~\eqref{eq:mu1}
  or~\eqref{eq:mu2}. 
\item Calculate $|\chi^{(k)}_{i,\rm out}(t_N;t_N)\rangle$ from
  Eq.~\eqref{chitj} and  
  propagate it backwards using $\hat{H}$
  from $t=t_N$ to the previous splitting time $t_{N-1}$. The resulting
  state is $|\chi^{(k)}_{i,\rm in}(t_{N-1})\rangle$.
\item\label{item:glue} At $t=t_{N-1}$, calculate $|\chi^{(k)}_{i,\rm
    out}(t_{N-1};t_{N-1})\rangle$ 
  from Eq.~\eqref{chitj}   
  and 'glue' to obtain 
  $|{\chi}^{(k)}_i(t_{N-1})\rangle =  
  |\chi^{(k)}_{k,\rm in}(t_{N-1})\rangle
  + |\chi^{(k)}_{i,\rm out}(t_{N-1};t_{N-1})\rangle$. This procedure is performed in
  the CIS basis for each channel wave function.
\item\label{item:bprop} Propagate $|{\chi}^{(k)}_i(t_{N-1})\rangle$ from
  $t=t_{N-1}$ to $t_{N-2}$ using $\hat{H}$ to obtain
  $|\chi^{(k)}_{i,\rm in}(t_{N-2})\rangle$. 
\item Repeat steps~\eqref{item:glue} and~\eqref{item:bprop} for all
  remaining  splitting times and propagate backward 
  up to $t=0$. During the backward propagation, the resulting wavefunction is
  stored in the CIS basis. As previously detailed, this procedure allows for
  performing the ``glueing'' procedure only once at every splitting time. It
  gives gives rises to the first term in Eq.~\eqref{eq:chi_total}. The second
  term involving the evaluation of
  the outer part (coherent summation) at any arbitrary time $t$ is obtained upon application
  Eq.~\eqref{eq:tilde2} to each of the individual contribution $|\chi_{i,\rm out}(t_j;t_j)\rangle$
  for all splitting times.
\end{enumerate}
\item\label{item:forw} Forward propagation and update of control: 
\begin{enumerate}
\item Determine the zeroth order approximation of the new control
  at times $(n+1/2)\Delta t$,   $\mathbf{A}_0^{(k+1)}(n+1/2\Delta t)$, 
  from  Eq.~\eqref{eq:mykrotov}, using the states at  times
  $n\Delta t$, i.e., the co-state obtained in step~\ref{item:backw},
  $|\chi^{(k)}_i(n\Delta t)\rangle$
  and  $|\varphi^{(k+1)}_i(n
  \Delta t)\rangle$  obtained with the control  
  $\mathbf{A}^{(k+1)}((n-
  1/2)\Delta t)$.  
\item If $\lambda_\omega\neq 0$ or  $\lambda_e\neq 0$,  solve
  Eq.~\eqref{transfer} to obtain $\tilde{\mathbf{A}}^{(k+1)}(\omega)$, 
  using the approximated $\mathbf{A}^{(k+1)}_0(t)$, and Fourier transform
  $\tilde{\mathbf{A}}^{(k+1)}(\omega)$ to time domain. 
\end{enumerate}
\item Increase $k$ by one and repeat steps~\ref{item:backw}
  and~\ref{item:forw} until convergence  
  of $J_T$ is reached.
\end{enumerate}
At this point, we would like to stress that the
parameters chosen for the momentum grid require 
particular attention for the optimization algorithm to work. This is due to 
the transformation from the CIS representation  to the Volkov basis
(CIS--to--$p$ transformation) at each splitting time, 
as discussed in Section~\ref{subsec:wfsm}. 
During the backward propagation, correspondingly, 
the inverse transformation is required, i.e., the $p$--to--CIS
transformation. The CIS--to--$p$ transformation of the outer part is 
evaluated using Eq.~\eqref{eq:Ci;tj}; the inverse of this
transformation is straightforwardly derived. 
Since the dynamics is reversible, forward propagation
(involving wavefunction splitting and the CIS--to--$p$ transformation)
needs to give identical results to  
backward propagation (involving wavefunction ``glueing'' and the
$p$--to--CIS transformation). 
This can and needs to be used to check the numerical accuracy of the
CIS--to--$p$ transformation and its inverse:
Since the inverse transformation involves integration over $\mathbf{p}$, 
a significant error is introduced if the sampling of the momentum grid
is  insufficient. 
Consequently, transforming the outer part from the CIS representation
to the Volkov basis and then back may not yield exactly
the same wave function. 
While for each $p$--to--CIS transformation
the error may be relatively small, it accumulates as the optimization proceeds 
iteratively according to Eq.~\eqref{eq:mykrotov}. It results in 
optimized pulses with non-physical and undesirable ``jumps'' at those
splitting times where the accuracy of the $p$--to--CIS 
transformation is insufficient and destroys the monotonic convergence
of the optimization algorithm. The jumps disappear when 
the number of the momentum grid points is increased and  $p_{max}$
is adjusted. 

Therefore, a naive solution to this problem 
would  be to considerably enlarge the number of
momentum grid points. However, this will significantly increase the
numerical effort of the optimization, i.e., evaluation of
the inner product in the rhs. of Eq.~\eqref{eq:overlap}.
The inner product involves not only calculation of the overlap of the inner
part in the CIS representation and  the outer part in the Volkov
basis but it also requires evaluation of the mixed terms,  
$\langle\chi^{(k)}_{i, \rm in}(t)|\hat{p}_z|\varphi^{(k+1)}_{i, \rm
out}(t)\rangle$ and $\langle\chi^{(k)}_{i, \rm out}(t)|\hat{p}_z|\varphi^{(k+1)}_{i, \rm
in}(t)\rangle$ and thus one CIS--to--$p$ transformation and
integration over two---perhaps even three---degrees of freedom
at every time $t$, for each channel $i$ and in every iteration step
$k+1$. Thence, finding the best balance between efficiency and
accuracy in the   $p$--to--CIS transformation is 
essential for the proper functioning and feasibility of the
optimization calculations. Also, reducing the total size of the radial
coordinate while simultaneously increasing the number of splitting times
translates into a more important number of evaluations of the inner product defined in
Eq.~\eqref{eq:overlap} in momentum representation. Below, we state explicitly the momentum
grid parameters utilized in our simulations which allowed 
for a good compromise between efficiency and accuracy.

\section{Application I: Prescribing the complete 
  photoelectron distribution} 
\label{sec:appl1}

We consider, as a first  example, the optimization of the complete
photoelectron distribution,
cf. Eq.~\eqref{JT1}, for a hydrogen atom. 
The wavepacket is represented, according to Eq.~\eqref{wavefunction}, in 
terms of the ground state $|\Phi_0\rangle$ and 
excitations $|\Phi^a\rangle$. The 
calculations employed a pseudo-spectral grid with
density parameter $\zeta=0.50$~\cite{GreenmanPRA10}, a  
spatial extension of $200\,$a.u. 
and 800 grid points. All optimization calculations employed a
linearly polarized electric field  
along the $z$ axis. This translates into a rotational symmetry of the photoelectron
distribution along the $z$ axis. Therefore, only wave functions of the form 
$\Psi_{\rm out}=\Psi_{\rm out}(p,\theta)$ need to be 
considered. For the calculation of the spectral components, the 
outer parts of the wave functions were  
projected onto the Volkov basis, defined 
on a spherical grid in momentum representation $\mathbf{p}$. For our calculations,
we adopted an evenly spaced grid in $p$ as well as in the polar coordinate
$\theta$.
The size of the radial component of the spherical momentum grid was set  
to $E_{max}=6\,$a.u., sampled at $301$ points. The same number of
points was utilized for the polar coordinate.
The splitting radius was set to $r_c=50\,$a.u., the total number of
splitting times is $N=3$ with a smoothing  
parameter $\Delta=5.0\,$a.u.~\cite{AntoniaPRA14}. The splitting procedure was applied
every $30\,$a.u. of time. Finally, a total integration time of $120\,$a.u. 
with a time step of $0.05\,$a.u. was utilized for the
time propagation. 

\begin{figure}[tb]
\centering
\includegraphics[width=0.95\linewidth]{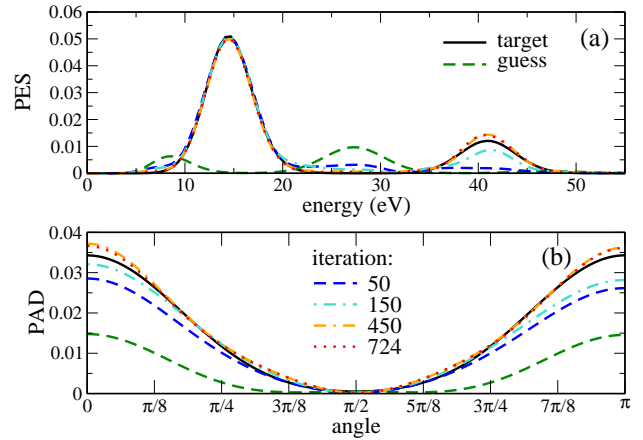}
  \caption{Optimal control of the complete photoelectron distribution
    for a hydrogen atom: (a) angle-integrated PES, 
    and (b) energy-integrated PAD.
    As the optimization 
    proceeds iteratively, the actual photoelectron distribution
    approaches the desired one (black 
    solid line) in both its energy dependence and angular
    distribution. The photoelectron distribution
    obtained with the guess field (green          
    dashed lines) is far from the desired distribution.
  }
\label{fig:func1_iterations}
\end{figure}
\begin{figure}[tb]
\centering
\includegraphics[width=0.95\linewidth]{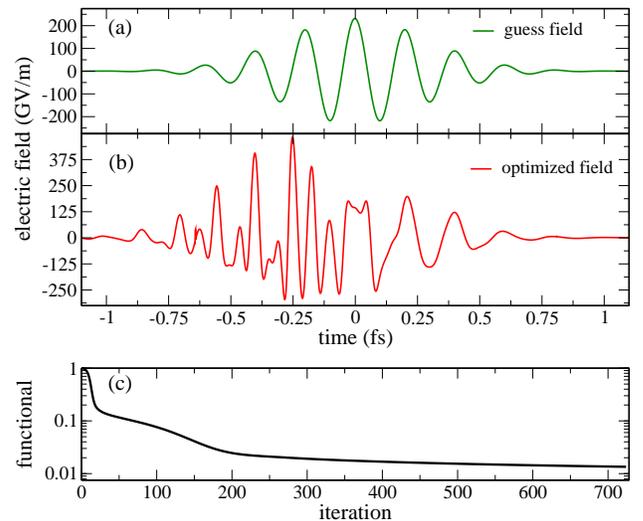}
\caption{Optimization of the full photoelectron distribution:
  (a) Guess field $E^{(0)}_z(t)$  chosen to start the optimization
  shown in Fig.~\ref{fig:func1_iterations} 
  and (b) optimized electric field obtained after 
  about 700 iterations. (c) The final time cost functional
  $J^{(1)}_T$  decreases monotonically, as
  expected for Krotov's method.}
\label{fig:func1_combined_1}
\end{figure}
We consider first the
minimization of the functional $J^{(1)}[\varphi,\varphi^\dagger]$ defined in 
Eq.~\eqref{JT1}. The goal is to find a vector potential ${A}_{z,\rm opt}(t)$ 
such that the  photoelectron distribution resulting from the electron
dynamics generated by ${A}_{z,\rm opt}(t)$ coincides with
$\sigma_0(\mathbf{p})$ at every point $\mathbf{p}$, cf. Eq.~\eqref{JT1}. For 
visualization convenience, we plot the angle-integrated PES and energy-integrated PAD, 
cf. Eq.~\eqref{eq:PAD_PES}, associated to ``target'' photoelectron
distribution $\sigma_0(\mathbf{p})$, as shown by the solid-black lines  
in Figs.~\ref{fig:func1_iterations}(a) and~\ref{fig:func1_iterations}(b),
respectively. To
simplify the optimization, neither frequency restriction nor amplitude
constraint on ${E}_{z}(t)$ is imposed, i.e.,
$\lambda_\omega=\lambda_e=0$. The initial guess for the vector
potential is chosen in such a way that the fidelity with respect 
to the target $\sigma_0(\mathbf{p})$ is poor, see the green dashed lines in 
Fig.~\ref{fig:func1_iterations}. 
Despite the bad
initial guess, the optimization quickly approaches the desired
photoelectron distribution,
converging monotonically, as expected for Krotov's method and
demonstrated in Fig.~\ref{fig:func1_combined_1}(c): After about 700 
iterations, the target distribution is realized with an error of 2\%. 
The reason for such a large number of iterations can be
understood by considering that the optimized photoelectron
distribution must coincide (point-by-point) with a two-dimensional target
object. This represents a non-trivial optimization problem. 
The optimized electric field is shown in
Fig.~\ref{fig:func1_combined_1}(b): Compared to the initial guess,
cf. Fig.~\ref{fig:func1_combined_1}(a), the amplitude of the optimized
field is somewhat increased, and a high-frequency oscillation has been
added. The monotonic convergence towards the target distribution  
in terms of angle-integrated PES and energy-integrated PAD 
is illustrated in  Fig.~\ref{fig:func1_iterations}. We can
appreciate that the algorithm first tends to match all points with higher
values, starting with the peak near 15$\,$eV, while adjusting the
remainder of the spectrum, with lower values, later in the
optimization. The slow-down of convergence, observed  in
Fig.~\ref{fig:func1_combined_1}(c) after about 200 iterations, is
typical for optimization methods that rely on gradient information
alone: As the optimum is approached, the gradient 
vanishes~\cite{EitanPRA11}. Such a
slow-down of convergence can only be avoided by incorporating
information from higher order derivatives in the
optimization. This is rather non-trivial in the framework of Krotov's
method~\cite{EitanPRA11,JaegerPRA14} and beyond the scope of our
current study.   

\section{Application II: Minimizing the probability 
  of emission into the upper hemisphere}
\label{sec:appl2}

As a second application of our control toolbox, we are interested in
minimizing the probability of emission into the upper hemisphere
without imposing any specific constraint on the number
of electrons emitted into the lower hemisphere. The final time cost
functional is given by Eq.~\eqref{JT2} with $\lambda_{2}^{+} > 0$  
and $\lambda_{2}^{-} = 0$. We consider again a hydrogen atom and a
linearly polarized electric field along the $z$-axis, using
the same numerical parameters as in Section~\ref{sec:appl1}.

\begin{figure}[tb]
\centering
\includegraphics[width=0.95\linewidth]{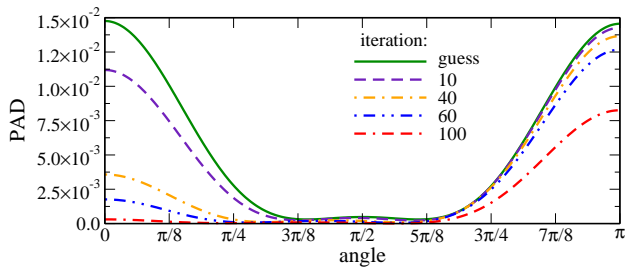}
\caption{Minimizing, for a hydrogen atom, photoelectron emission
  into the upper hemisphere: As the optimization proceeds iteratively, 
  the probability of emission into the upper 
  hemisphere decreases monotonically up to almost complete extinction.
}.
\label{fig:appl2:angle}
\end{figure}
\begin{figure}[tb]
\centering
\includegraphics[width=0.95\linewidth]{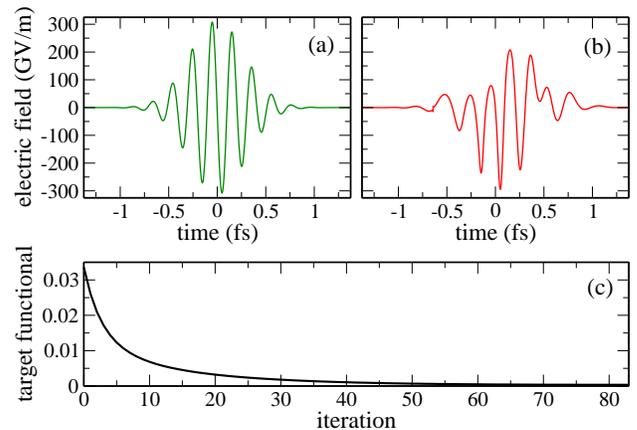}
\caption{Minimization of the probability of emission into the upper hemisphere for
  hydrogen: Guess (a)  and optimized (b) electric field
  for the optimization shown in Fig.~\ref{fig:appl2:angle}. Also for
  this target functional, Eq.~\eqref{JT2},  
  monotonic convergence of the optimization algorithm is achieved (c).} 
\label{fig:combined_zero1}
\end{figure}
In contrast to the example discussed in Section~\ref{sec:appl1}, no 
particular expression for the target  PES and PAD
needs to be imposed---we only require the probability of emission into
the upper hemisphere to be minimized regardless of the actual  
shape of angle-integrated PES and energy-integrated PAD. We employ the
optimization prescription 
described in Section~\ref{subsec:krotov} using 
Eq.~\eqref{eq:mu2} in the final time condition for the adjoint state. As the optimization
proceeds iteratively, the energy-integrated PAD becomes more and more asymmetric, see
Fig.~\ref{fig:appl2:angle}, minimizing emission into the upper hemisphere, as desired. 
The guess and optimized pulses are
shown in Fig.~\ref{fig:combined_zero1}(a) and~(b). 
As illustrated by the solid green line in Fig.~\ref{fig:appl2:angle}, 
the guess field was chosen such that it leads to a 
symmetric probability of emission for the two hemispheres. 
Again, monotonic convergence of the final time cost functional is
achieved, cf. Fig.~\ref{fig:combined_zero1}(c). At the end of the iteration
procedure, the probability of emission into the upper hemisphere vanishes
completely. 
As for the lower hemisphere, the emission probability initially remains almost
invariant as the algorithm proceeds iteratively, see
Fig.~\ref{fig:appl2:angle},  while the probability of emission into
the upper hemisphere decreases very fast, and  
monotonically, as expected. However, for a large
number of iterations, the probability of emission into the lower 
hemisphere starts to decrease as well. After about 150 iterations it
reaches an emission probability of $2.3\times 10^{-4}$, that is
two orders of magnitude smaller than for the guess pulse.
Although our goal is only for the probability of emission into the upper 
hemisphere to be minimized, without specific constraints on the
probability of emission into the lower hemisphere, the current results
are completely consistent in terms of the 
optimization problem. More precisely, the optimization does exactly
what the functional $J^{(2)}_T$, Eq.~\eqref{JT2} with 
$\lambda_{2}^{+} > 0$  and $\lambda_{2}^{-} = 0$, targets. In fact,
since the target  functional depends on the 
upper hemisphere alone, then, by construction, the algorithm calculates the
corrections to the field according to Eq.~\eqref{eq:mykrotov}, 
regardless of how these changes affect the probability of emission into the
lower hemisphere. To keep the probability of emission into the lower 
hemisphere constant or to maximize it, an additional optimization functional
is required. This is investigated in the following section and defines the
motivation for the maximization of the anisotropy of emission discussed in the following
lines.

\section{Application III: Maximizing the difference in 
  the number of electrons emitted 
  into upper and lower hemisphere}
\label{sec:appl3}

Finally we maximize 
the difference in probability for emission into the upper and the lower
hemispheres. To this end, we construct the final-time cost functional
such that it maximizes emission
into the upper hemisphere while simultaneously  minimizing emission
into the lower hemisphere.
This is expressed by the functional~\eqref{JT2} where
both weights are non-zero and have different signs, 
$\lambda_2^{(+)} < 0$ and  $\lambda_2^{(-)} > 0$. The signs
correspond to maximization and minimization, 
respectively. We consider this control
problem for two different atoms---hydrogen as a one-channel case and
argon as an example with three active channels~\cite{AntoniaPRA14}. 
The latter serves to underline the appropriateness of our methodology for quantum control
of multi-channel problems. 

Furthermore, in order to demonstrate the versatility of our optimal
control toolbox in constraining specific properties of the optimized electric
field, we consider the following options: (i) a spectral constraint,
i.e., $\lambda_\omega\neq 0$ in Eq.~\eqref{eq:Comega}, and (ii) the
constraint to minimize fast changes in the vector potential, with 
$\lambda_e\neq 0$ in  Eq.~\eqref{eq:Ce}. The latter is equivalent to
avoiding large electric field amplitudes. 

\subsection{Hydrogen}
\label{subsec:appl3:h}

\begin{figure}[tb]
\centering
\includegraphics[width=0.95\linewidth]{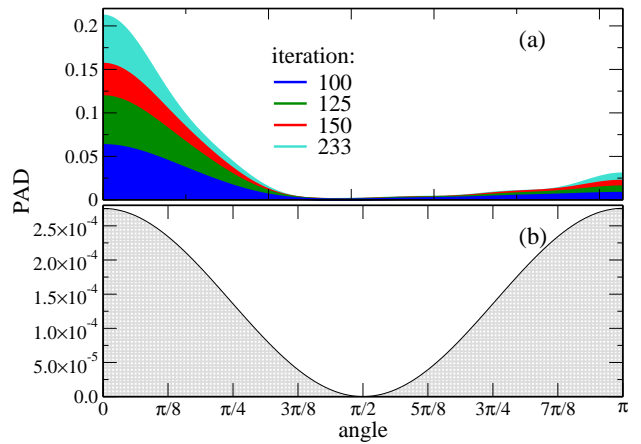}
\caption{Maximizing, for a hydrogen atom, the difference in
    photoelectron emission into the upper and lower hemisphere: (a) The
    probability for emission into the upper hemisphere
    ($0\le\theta\le\pi/2$) increases significantly as
    the optimization proceeds. Although the probability for
    emission with angles ($\pi/2\le\theta\le\pi$) also grows somewhat,
    the overall difference increases. The energy-integrated PAD
    obtained with the guess 
    pulse is shown in (b). Note 
    the different y-axis scales in (a) and (b). 
}
\label{fig:H_pes}
\end{figure}
\begin{figure}[tb]
  \centering
  \includegraphics[width=0.95\linewidth]{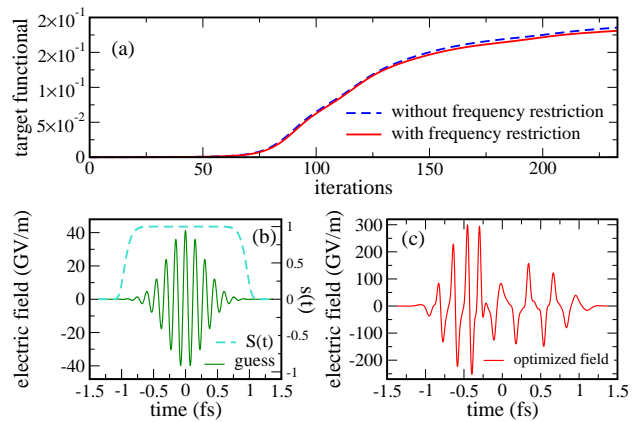}
  \caption{Maximization of the anisotropy in the PAD for hydrogen: The
    target functional $J_T^{(2)}$, for the
    optimization shown in Fig.~\ref{fig:H_pes}, 
    measuring the difference in
    probability for emission into upper and lower hemisphere increases
    monotonically with ($\lambda_\omega\neq0$) and without
    ($\lambda_\omega=0$) spectral constraint (a). The guess field (green
    line) is shown in (b) together with the shape function $s(t)$ used
    in both optimizations. The optimized field obtained with the
    spectral constraint is displayed in (c). 
}
\label{fig:H_bdw1}
\end{figure}
\begin{figure}[tb]
\centering
\includegraphics[width=0.95\linewidth]{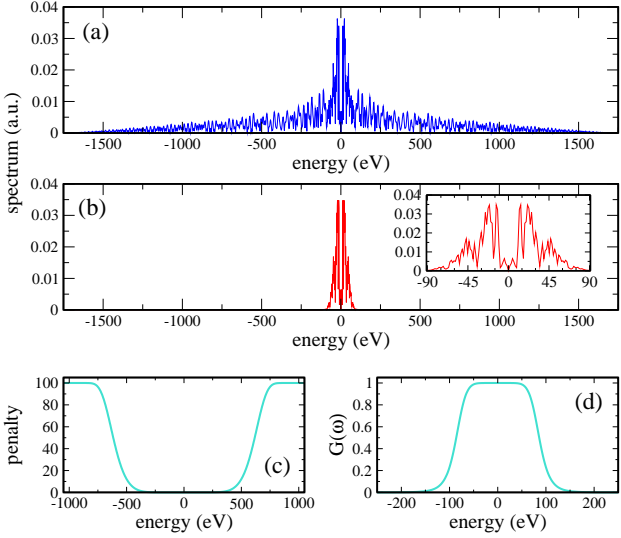}
\caption{Hydrogen, maximization of the anisotropy of emission:
    Spectrum of the optimized electric field for the
    optimization shown in Figs.~\ref{fig:H_pes} and~\ref{fig:H_bdw1}
    with (b) and without (a)
    spectral constraint. The corresponding penalty function
    $\tilde\gamma(\omega)$ and transfer function $\tilde{G}(\omega)$,
    cf. Eqs.~\eqref{eq:Comega} and~\eqref{eq:Gw}, 
    are shown in (c) and (d), respectively.
}
\label{fig:frequency_components}
\end{figure}

We consider a hydrogen atom, interacting with an electric field
linearly polarized along the $z$-axis, using the same numerical parameters as
in Sec.~\ref{sec:appl1}. The optimization was carried out 
with and without restricting the spectral bandwidth of $E_z(t)$. 
Figure~\ref{fig:H_pes}(b) displays the symmetric energy-integrated PAD
obtained with the  
Gaussian guess field, shown in Fig.~\ref{fig:H_bdw1}(b), for which a
central frequency $\omega_0=27.2\,$eV was used. For the optimization
with spectral constraint, 
the admissible frequency components for $E_z(t)$ 
are chosen such that $\big|E_z(\omega)\big|^2 \le\epsilon
\,\hspace{0.1cm} \text{for all}\hspace{0.1cm} |\omega| \ge \omega_{max}$ with 
$\omega_{max}=5\,$a.u.$\,\approx 136.1 \,$eV. 
This requirement translates into the penalty function
$\tilde{\gamma}(\omega)$ shown in Fig.~\ref{fig:frequency_components}(c),
for which we have used the form
\begin{eqnarray}
  \label{eq:filterfunc}
  \tilde{\gamma}(\omega) &=& \tilde{\gamma}_0\, \left(1-e^{-(\omega/\alpha)^{2n}}\right),
\end{eqnarray}
where the parameters $\alpha$, $n$ and $\tilde{\gamma}_0$ must be chosen such that 
the term  $\lambda_\omega\,\hat{\gamma}(\omega)$  in
the functional $C_{\omega}[\mathbf{A}]$ in Eq.~\eqref{eq:Comega} takes very large values in the region
of undesired frequencies. For our first example, $\alpha=25$, $n=6$ and $\tilde{\gamma}_0=1$
allows for strongly penalizing, and therefore filtering all undesirable frequency
components above $|\omega|\ge\omega_{max}$, as it is shown by the corresponding
transfer function $\tilde{G}(\omega)$, cf. Fig.~\ref{fig:frequency_components}(d). 
Note that it is not the 
weight $\lambda_{\omega}$ alone that determines how strictly the
spectral constraint is enforced; it is the ratio $\lambda_{\omega}/\lambda_a$
that enters in the transfer function $\tilde{G}(\omega)$. This reflects the
competition of the different terms in the complete optimization
functional, Eq.~\eqref{eq:Jtotal}. 

As in the previous two examples, the optimization approach  developed
leads to monotonic convergence of the
target functional, Eq.~\eqref{JT2}, with and without spectral
constraint. This is illustrated in Fig.~\ref{fig:H_bdw1}(a). Even
though the spectra of the fields optimized with and without spectral
constraint, are completely different,
cf. Fig.~\ref{fig:frequency_components}(a) and~(b), the speed of convergence is
roughly the same, and the maximum values for $J_T^{(2)}$ reached using both fields
are also very
similar, cf. Fig.~\ref{fig:H_bdw1}(a). This means that the algorithm finds two
distinct solutions. Such a finding is very encouraging
as it implies that the spectral
constraint does not put a large restriction onto the control
problem. In other words, more than one, and probably many, control
solutions exist, and it is just a matter of picking the suitable
one with the help of the additional constraint.  It also implies that
most of the frequency components in the spectrum of the field
optimized without spectral constraint are probably not essential.  
This is verified by removing the undesired spectral components in
Fig.~\ref{fig:frequency_components}(a), using the same transfer function
utilized for the frequency-constrained optimization shown in
Fig.~\ref{fig:frequency_components}(d). 
The energy-integrated PAD obtained with
such a filtered optimized pulse remains asymmetric, and the value of
the target functional $J_T^{(2)}$ is decreased by only about 10 per
cent. 

The peak amplitude of the optimized field is about one order of
magnitude larger than that of the guess field, cf. 
Fig.~\ref{fig:H_bdw1}(b) and~(c). 
The increase in peak amplitude is
connected to the gain in emission probability for the northern
hemisphere by almost three orders of magnitude. The optimized pulse
thus ionizes much more efficiently than the guess pulse.
\begin{figure}[tb]
\centering
\includegraphics[width=0.95\linewidth]{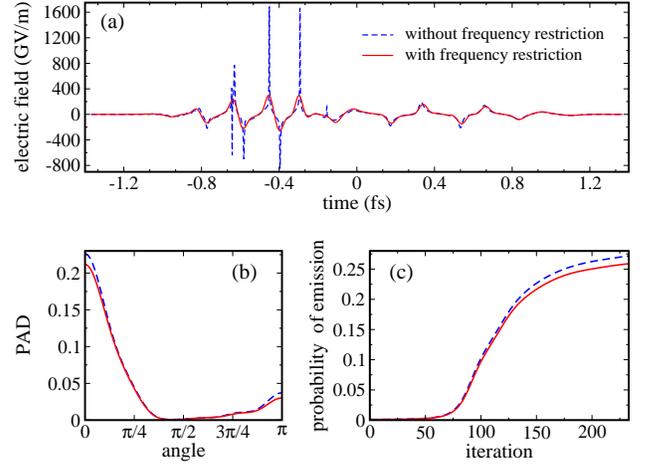}
\caption{Maximization of the difference of photoelectrons emitted into the lower
  and upper hemispheres for hydrogen: Optimized electric fields with
  ($\lambda_{\omega}\neq0$) and without ($\lambda_{\omega}=0$)
  frequency restriction (a) where the red curve shows the same data
  as in Fig.~\ref{fig:H_bdw1}(c). Also compared are 
  the energy-integrated PAD  (b) and total emission
  probability (c) obtained with 
  the frequency-constrained and unconstrained optimized 
  fields.}
\label{fig:H_contrast}
\end{figure}
Figure~\ref{fig:H_contrast}(a) compares the electric fields  optimized
with and without spectral constraints---a huge difference is observed
for the two fields. While the  electric field  optimized without
spectral constraint presents very sharp and high peaks 
in amplitude, beyond experimental feasibility, 
the frequency-constrained optimized 
field is characterized by reasonable amplitudes and a much smoother
shape. The frequency components of the unconstrained field shown in 
Fig.~\ref{fig:frequency_components}(a) now become clear.
Note that the difference in amplitude only
appears during the first half of the  overall pulse duration, see 
Fig.~\ref{fig:H_contrast}(a). It is a known feature of Krotov's method to favor 
changes in the field in an asymmetric fashion; the feature results from the 
sequential update of the control, as opposed to a concurrent
one~\cite{SophieNJP2011}. 

Figure~\ref{fig:H_contrast}(b) shows the energy-integrated PAD 
obtained  upon propagation with the two fields. One notes 
that, although the probability of emission into the lower hemisphere is 
larger for the unconstrained than for the constrained field, the
same applies to the probability of emission into the upper 
hemisphere. Therefore the
difference in the number of electrons emitted into upper and lower
hemisphere is in the end
relatively close, which explains the behavior of the
final-time functional observed in 
Fig.~\ref{fig:H_bdw1}(a). The electron dynamics generated by the
frequency-unconstrained field leads to a larger total probability of
emission into both hemispheres, with respect to that obtained with the
frequency-constrained field, as shown in  
Fig.~\ref{fig:H_contrast}(c). More precisely, propagation with the 
unconstrained optimized field results in a total probability of emission of
$0.27$, i.e., probabilities of  $0.23$ and $4.3\times
10^{-2}$ for emission into the upper and lower hemisphere,
respectively. In comparison, a total probability of emission
of $0.26$ is obtained for the frequency-constrained field, with
probabilities of  
emission into the upper and lower hemispheres of $0.22$ and
$3.9\times 10^{-2}$, respectively.                     
The fact that the spikes observed in the unconstrained optimized field 
do not have any significant impact on the asymmetry of the PAD
 can be rationalized by the short timescale on which the intensity is
 very high. This time is too short for the 
electronic system to respond to the rapid variations of the field
amplitude. 

\begin{figure}[tb]
\centering
\includegraphics[width=0.95\linewidth]{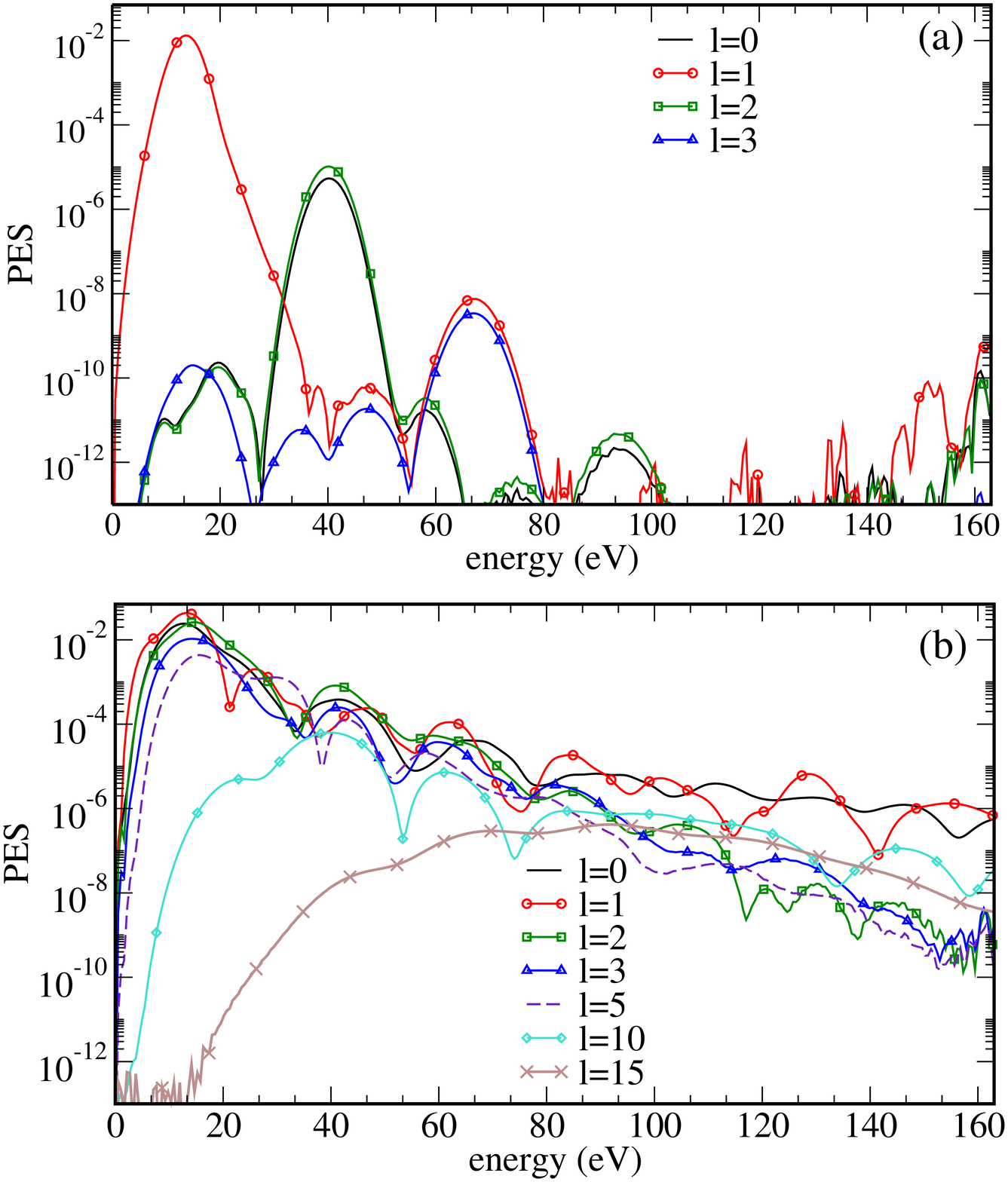}
\caption{Maximizing  the anisotropy of photoelectron emission for
  hydrogen: Partial wave  
  contribution to the angle-integrated PES, shown in
  Fig.~\ref{fig:H_pes}, 
  obtained with the guess (a) and the frequency-constrained
  optimized field (b).
}
\label{fig:anto2_v2}
\end{figure}

In order to rationalize how anisotropy of electron emission is
achieved by the optimized field, we analyze in Fig.~\ref{fig:anto2_v2} the partial
wave decomposition of the angle-integrated PES, comparing the results
obtained with the guess field to those obtained with the
frequency-constrained optimized field. Inspection of Fig.~\ref{fig:anto2_v2} reveals
that upon optimization, there is a clear transition from distinct ATI peaks, Fig.~\ref{fig:anto2_v2}(a),
to a quasicontinuum energy spectrum, Fig.~\ref{fig:anto2_v2}(b). 
Also, the optimized field enhances the contribution of states of higher
angular momentum that have the same kinetic energy.  In particular,
the peaks for $l=5$ are dramatically higher than in the PES obtained
with the guess field.  In fact, the symmetric case, cf. 
Fig.~\ref{fig:anto2_v2}(a), shows an energy  distribution of partial waves characterized by
waves of the same parity at the same energy, whereas the asymmetric case reveals
a partial wave distribution of opposite parity at the same energy, cf. Fig.~\ref{fig:anto2_v2}(b). 
Figure~\ref{fig:anto2_v2} thus demonstrates that the desired
asymmetry in the energy-integrated PAD is achieved through the mixing
of various partial waves of opposite parity at the same energy.
Interestingly, 
especially lower frequencies are mixed with a considerable intensity into the pulse spectrum which 
leads to higher order multiphoton ionization leading to comparable final energies in the PES. 
Thus, more angular momentum states are mixed.

\begin{figure}[tb]
\centering
\includegraphics[width=0.95\linewidth]{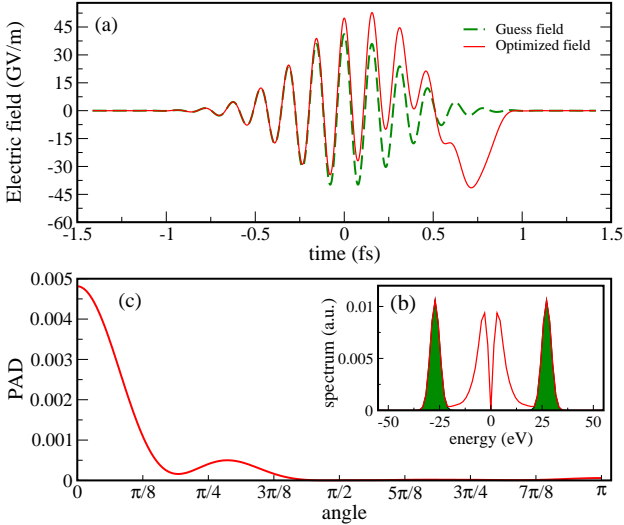}
\caption{Maximizing  the anisotropy of photoelectron emission for hydrogen:
  Optimization results obtained when simultaneously constraining the maximal
  amplitude and frequency components of the electric field for the weights
  $|\lambda^{(-)}_{eff}|=2|\lambda^{(+)}_{eff}|$ with $|\lambda^{(+)}_{eff}|=1$. A perfect anisotropy of
  emission is obtained.
}
\label{fig:newH1}
\end{figure}

Next, we would like to constrain not only the frequency components
but also the maximal field amplitude, as the maximal field amplitude of the 
electric field, shown in Fig.~\ref{fig:H_bdw1}(c) is still important.
To this end, we employ Eq.~\eqref{eq:newmethod} for $\lambda_e > 0$, which
penalizes large changes on the derivative of the vector potential, cf.
Eq.~\eqref{eq:Ce}, and thus large values of the electric 
field amplitude. As can be seen in
Fig.~\ref{fig:newH1}(a), the resulting optimized field is one order of magnitude
smaller than that for which no amplitude restriction was imposed, cf.
Fig.~\ref{fig:H_bdw1}(c), and of the same order of magnitude as the
guess field. 
Despite the constraint and 
as shown in Fig.~\ref{fig:newH1}(c), a perfect top-bottom 
asymmetry is obtained. 

A common feature observed between the amplitude-unconstrained and
constrained cases concerns the low frequencies appearing upon
optimization, cf. Fig.~\ref{fig:frequency_components}(b)  
and Fig.~\ref{fig:newH1}(b), respectively. 
To quantify the role of the frequency components for achieving
anisotropy, we start by 
suppressing all frequency components above                               
$10$ eV: the
anisotropy of emission is preserved. On the other hand, removing frequencies
below the XUV                                                 
re-establish the initial symmetry of emission into both
hemispheres. 
Therefore, in both cases  the top-bottom asymmetry arises
from low frequency components of the optimized field 
and is achieved through the mixing
of various partial waves of opposite parity at the same energy.

\subsection{Argon}
\label{subsec:appl3:ar}

We extend now our quantum control 
multi-channel approach to the study of electron dynamics in argon, 
interacting with an electric field linearly polarized along the 
$z$-direction. We consider the $3s$ and $3p$ orbitals to contribute 
to the ionization dynamics and define three ionization channels $3s$, $3p$ with
$m=0$ and $3p$ with $m=+1$ (the case $3p$ with $m=-1$ is symmetric
to $m=+1$ due to the polarization direction of the electric field,
linearly polarized along to the $z$ axis). 
In order to describe the multi-channel dynamics, a
spatial grid of $100\,$a.u. with $450$ grid points and a density
parameter of $\zeta=0.55$ was
utilized. The  
size of the radial component of the spherical momentum grid was set  
to $E_{max}=12\,$a.u., sampled by $601$ evenly spaced points, while
the polar component $\theta\in[0,\pi]$ was discretized using $301$
points. A splitting radius of $r_c=50\,$ a.u.,
and a smoothing parameter 
$\Delta=10.0\,$a.u. were employed, together with a splitting step of
$2.0\,$a.u. and 
a total number of $N_s=2036$ splitting times. For time propagation, the 
time step was chosen to be $0.01\,$a.u., for an 
overall integration time  $\Delta T\approx200\,$a.u.

\begin{figure}[tb]
\centering
\includegraphics[width=0.95\linewidth]{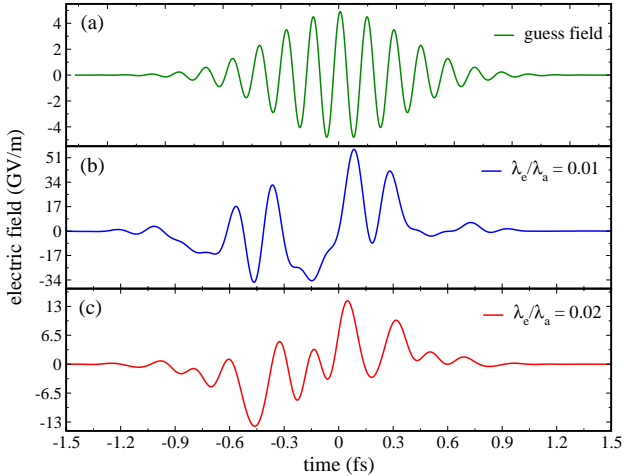}
\caption{Maximizing, for an argon atom, the difference in
  photoelectron emission into the upper and lower hemisphere:
  Guess field (a) utilized for the optimization. Optimized fields
  obtained with an amplitude constraint are depicted in (b) and (c) 
  respectively.                        
}
\label{fig:NewArgon1}
\end{figure}
\begin{figure}[tb]
\centering
\includegraphics[width=0.95\linewidth]{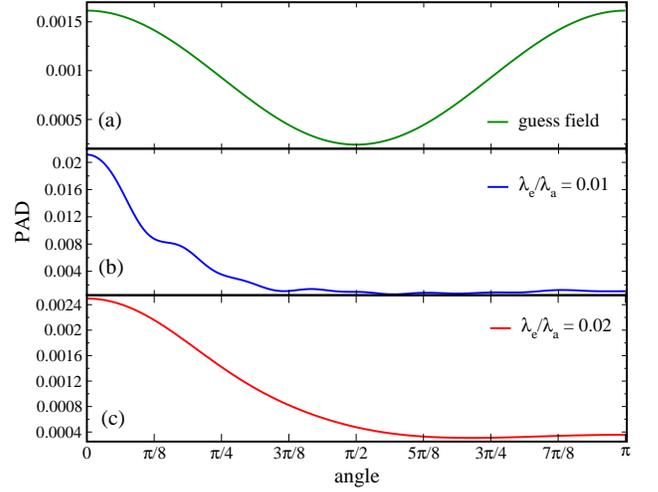}
\caption{Maximization of the top-bottom asymmetry in argon:
  Energy-integrated PAD obtained with the guess pulse
  (a) and amplitude-constrained cases with 
  $|\lambda^{(-)}_{eff}|=2|\lambda^{(+)}_{eff}|$ and 
  $|\lambda^{(+)}_{eff}|=1$
  in (b) and (c), respectively.   
  Note the different scales for 
  the probability of emission. 
}
\label{fig:NewArgon2}
\end{figure}
Analogously to the results shown for hydrogen in
Sec.~\ref{subsec:appl3:h}, the goal is to
maximize the difference in the probability for electron emission into
the upper and lower hemispheres. To start the optimization, a 
Gaussian-shaped guess electric field with central frequency
$\omega=27.2\,$eV and maximal amplitude $E_{max}=5.14\,$GV/m was
chosen. It is depicted in Fig.~\ref{fig:NewArgon1}(a) and  yields a
symmetric distribution for the upper  
and lower hemispheres, see 
Fig.~\ref{fig:NewArgon2}(a). The total emission probability amounts to
only $1.4\times 10^{-2}$. 
In order to obtain reasonable 
pulses which result in a maximally anisotropic PAD, 
we utilize Eq.~\eqref{eq:Ce} with $\lambda_e\neq0$
to minimize fast changes in the vector
potential and avoid large peaks of the electric field amplitude. 
 
\begin{figure}[tb]
\centering
\includegraphics[width=0.95\linewidth]{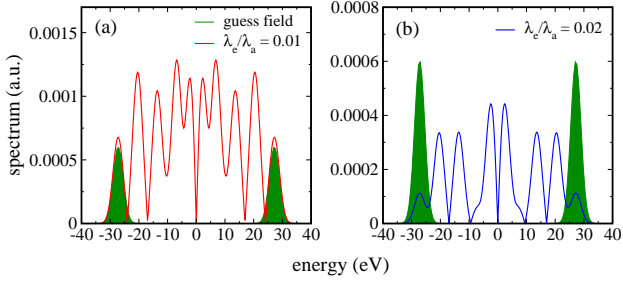}
\caption{Anisotropy of PAD in argon: Spectra of the optimized pulses 
  and the guess field for comparison. 
}
\label{fig:spectra_argon}
\end{figure}

The optimized pulses for two values of the ratio
$\lambda_e/\lambda_a$, charaterizing the relative weight of minimizing
peak values in the electric field compared to minimizing the
integrated vector potential, are shown in
Figs.~\ref{fig:NewArgon1}(b) and (c), respectively. As expected, a
larger amplitude constraint yields an electric field with a smaller
maximal amplitude. In fact, the maximal amplitude for
$\lambda_e/\lambda_a=0.01$  is one order of magnitude larger than that
of the guess field, whereas for $\lambda_e/\lambda_a=0.02$ 
it is only three times larger. 
Figures~\ref{fig:NewArgon2}(b) and (c) display the energy-integrated PADs
obtained with these fields. A significant top-bottom asymmetry of
emission is achieved in both 
cases, the main difference being the total emission probability of 
$2.7\times 10^{-2}$ for Fig.~\ref{fig:NewArgon2}(b) compared to
$9.4\times 10^{-3}$ for Fig.~\ref{fig:NewArgon2}(c). 
The spectra of the two optimized fields are examined in 
Fig.~\ref{fig:spectra_argon}. Despite the difference in 
amplitude, both optimized fields are characterized by
low frequency components. Note that no frequency restriction was
imposed. This finding suggests that the low frequency components are
responsible for achieving the top-bottom asymmetry. Indeed,  
removing all optical and infra-red (IR) components results in a
complete loss of the asymmetry. On the other hand, removing frequency
components above $10\,$eV does not affect the top-bottom asymmetry
achieved by both optimized fields considerably. 

\begin{figure}[tb]
\centering
\includegraphics[width=0.95\linewidth]{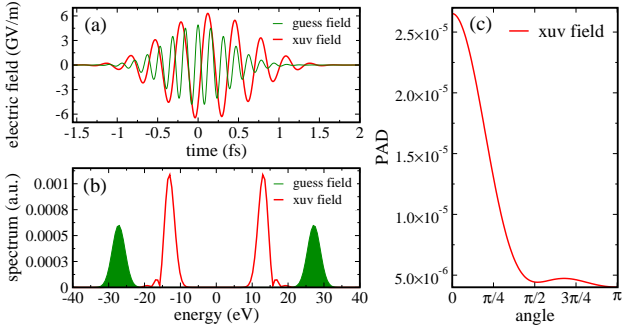}
\caption{Top-bottom asymmetry  in argon: Frequency and amplitude-constrained
  optimized field and its spectrum  in (a) and (b) with the guess   
  field shown for comparison. For obtaining the XUV field, the ratios
  $\lambda_e/\lambda_a = 0.02$ and $\lambda_{\omega}/\lambda_a=0.02$
  with a penalty function $\tilde{\gamma}_{XUV} = \tilde{\gamma}(\omega-\omega_0) 
  + \tilde{\gamma}(\omega+\omega_0)-1+\epsilon_{\omega} $
  with $\tilde{\gamma_o}=100$, $\omega_0 = 27.2$ eV, $n=4$ and $\alpha=15$ 
  were utilized, cf. Eq.~\eqref{eq:filterfunc}. The quantity
  $\epsilon_{\omega}=0.001$ has been introduced in order to avoid numerical instabilities
  when evaluating the transfer function $\tilde{G}(\omega)$, cf. Eq.~\eqref{eq:Gw}.
}
\label{fig:XUV_optim}
\end{figure}

These optimization results raise the question whether frequency
components in the optical and IR range 
are essential for achieving the top-bottom asymmetry or whether a pure
XUV field can also realize the desired control. To answer this
question, we now penalize all frequency components in the
optical and IR region. The resulting  optimized 
electric field                                    
and its spectrum  are depicted in Fig.~\ref{fig:XUV_optim}(a) and (b),
respectively.  This field indeed possesses frequency components only
in the XUV region, cf. Fig.~\ref{fig:XUV_optim}(b).
Nevertheless, a strongly asymmetric top-bottom emission is again
achieved, cf. Fig.~\ref{fig:XUV_optim}(c). Therefore, 
while optical or IR excitation may significantly contribute to 
achieving anisotropy of the photoelectron emission, fields with
frequency components in the XUV alone may also lead to such an
asymmetry.  

\begin{figure}[tb]
\centering
\includegraphics[width=0.95\linewidth]{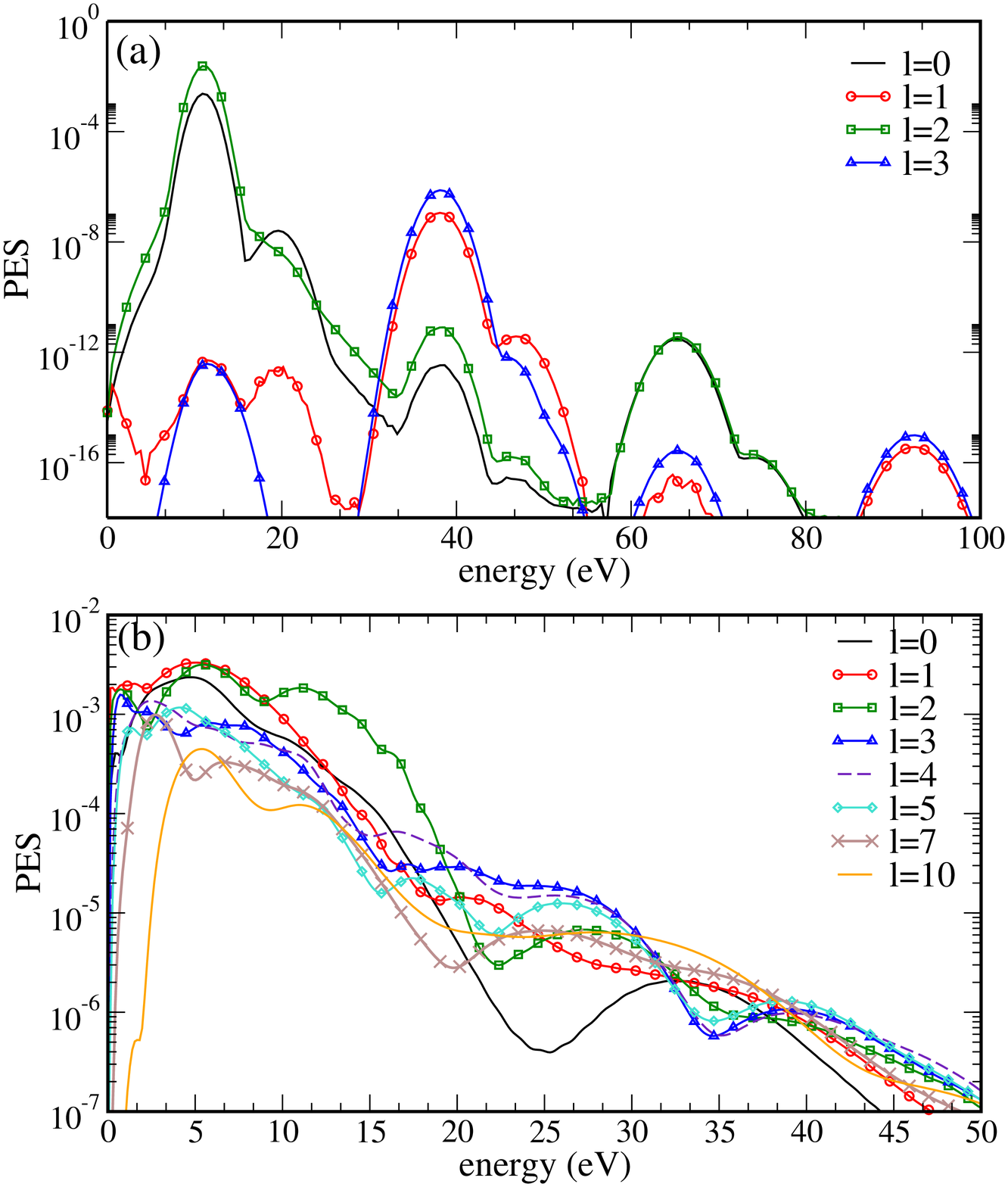}
\caption{Maximizing  the anisotropy of photoelectron emission for
  argon: Partial wave  
  contribution to the angle-integrated PES, shown in Fig.~\ref{fig:NewArgon2} 
  obtained with  the guess (a) and the amplitude-constrained optimized field
  corresponding to the ratio $\lambda_e/\lambda_a=0.01$ in (b). 
}
\label{fig:argon_partial}
\end{figure}

\begin{figure}[tb]
\centering
\includegraphics[width=0.95\linewidth]{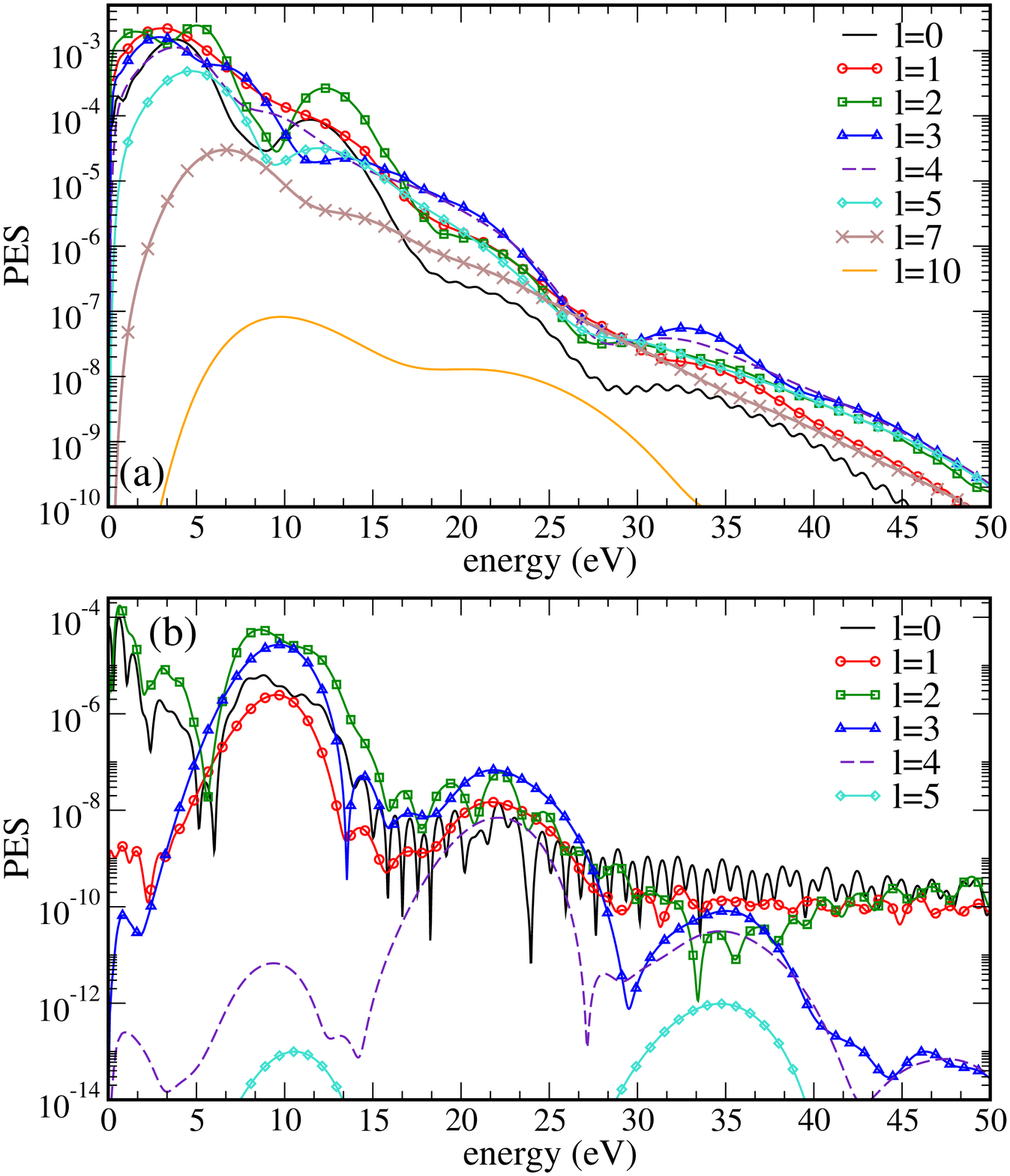}
\caption{Maximizing  the anisotropy of photoelectron emission for
  argon: Partial wave  
  contributions to the angle-integrated PES corresponding to the
  energy-integrated PAD shown in Fig.~\ref{fig:NewArgon2}(c) 
  obtained amplitude-constrained optimized field
  for the ratio $\lambda_e/\lambda_a=0.02$ and that corresponding to
  the PAD shown in Fig.~\ref{fig:XUV_optim}(c) obtained with the optimized XUV pulse in pannels (a) 
  and (b), respectively. 
}
\label{fig:argon_partial2}
\end{figure}

Finally, we would like
to understand the physical mechanism from which the anisotropy in the emission
into both hemispheres arises. To this end,  
we consider the partial wave decomposition of the angle-integrated
PES. 
Analogously to our analysis for hydrogen,
cf. Section~\ref{subsec:appl3:h}, a symmetric PAD,  as obtained with
the guess field, is characterized by an energy distribution of partial
waves of the same parity  at the same energy,
cf. Fig.~\ref{fig:argon_partial}(a). In contrast, the partial wave
decomposition  corresponding to the asymmetric PAD reveals an energy
distribution of partial waves of different parity at the same energy, 
cf. Figs.~\ref{fig:argon_partial}(b), \ref{fig:argon_partial2}(a).
For the optimized fields with significant optical and IR components,
many partial waves, including those with high angular momentum,
contribute to the angle-integrated PES. 
This suggests that the top-bottom anisotropy of photoelectron 
emission is achieved by absorbing low-energy photons at relatively high
intensity which is accompanied by strong mixing of a number of partial
waves of opposite parity at the same energy. The same
mechanism had previously been found for  
hydrogen, cf. Section~\ref{subsec:appl3:h}. 

As for the optimized XUV electric field yielding an asymmetric
probability for emission, shown in 
Fig.~\ref{fig:XUV_optim}(c), the same mechanism involving mixing of
partial waves of different parity 
at the same energy is found,
cf. Fig.~\ref{fig:argon_partial2}(b). Nevertheless, a much smaller
number of partial waves is involved,
cf. Fig.~\ref{fig:argon_partial2}(a) and (b).
For the XUV pulse (Fig.~\ref{fig:XUV_optim}(b)), it is  mainly the 
components of the continuum wavefunction with angular momentum $l=2$ and
$l=3$ that contribute to the anisotropy of emission. 

The reason why partial waves with different parity
are always present for anisotropic photoelectron emission 
can be straightforwardly understood. 
It lies in the fact that the angular distribution arises from products
of spherical harmonics, cf. Eqs.~\eqref{eq:PAD}, \eqref{eq:dE} and
\eqref{eq:Ci;tj},  and  the product of two spherical
harmonics with the same (opposite) parity is a symmetric
(antisymmetric) function of  $\theta$. 
The optimized pulses take advantage of this property and realize 
the desired asymmetry by driving the dynamics in such a way that it
results in partial wave components which interfere constructively
(destructively) in the upper (lower) hemisphere. 

We have also investigated whether channel coupling 
plays a role in the generation of the anisotropy.
While switching off the interchannel coupling in the dynamics under
the optimized pulse shown in Fig.~\ref{fig:XUV_optim}(a) decreases the
resulting anisotropy slightly, overall it still yields an anisotropic
PAD.  This shows that interchannel coupling in argon is not a key 
factor in achieving top-bottom asymmetry in photoelectron angular
distributions.

\section{Conclusions}

\label{sec:concl}


To summarize, we have developed a quantum optimal control toolbox to
target specific features in photoelectron spectra and photoelectron
angular distributions that result from the
interaction of a closed-shell atom with strong XUV radiation. 
To this end, we have combined Krotov's method for quantum
control~\cite{ReichJCP12} with the time-dependent configuration
interaction singles approach to treat the electron
dynamics~\cite{GreenmanPRA10} and the wave-function splitting 
method to calculate photoelectron
spectra~\cite{AntoniaPRA14,AntoniaPRA15Erratum}. 
We have presented here the algorithm and its implementation in detail. 
To the best of our
knowledge, our work is the first to directly target photoelectron
observables in quantum optimal control. 

We have utilized this toolbox to identify, for the benchmark systems
of hydrogen and argon atoms, photoionization pathways which result 
in asymmetric photoelectron emission. Our optimization results show that 
efficient mechanisms for achieving top-bottom asymmetry exist in both 
single-channel and multi-channel systems. We have found the channel
coupling to be beneficial, albeit not essential for achieving  asymmetric
photoelectron emission. Since typically the solution
to a quantum control problem is not unique, additional constraints are
useful to ensure certain desired properties of the control fields,
such as limits to peak amplitude and spectral width. We have
demonstrated how such constraints allow to determine solutions
characterized by low or high photon frequency. In the low frequency
regime, our control solutions require relatively high
intensities. Correspondingly, the anisotropy of the photoelectron
emission is realized by strong mixing of many partial waves. 
In contrast, for pure XUV pulses, we have found low to moderate peak
amplitudes to be sufficient for asymmetric photoelectron emission. In
both cases, we have identified the 
top-bottom asymmetry to originate from mixing, in the photoelectron
wavefunction, various partial waves of 
opposite parity at the same energy. The corresponding constructive
(destructive) interference pattern in the upper (lower) hemisphere
yields the desired asymmetry of photoelectron emission.
Whereas many partial waves contribute for control fields characterized
by low photon energy and high intensity, interference of two partial
waves is found to be sufficient in the pure XUV regime.
In all our examples, we have found surprisingly simple 
shapes of the optimized electric fields. In the case of 
hydrogen, tailored electric fields to achieve asymmetric
photoelectron emission have been discussed before and we can compare
our results to those of Refs.~\cite{ChelkowskiPRA2004,ShvetsovPRA14}.
Our work differs from these studies in that we avoid a parametrization
of the field and allow for complete freedom in the change the electric
field, whereas Refs.~\cite{ChelkowskiPRA2004,ShvetsovPRA14}
considered only the  carrier-envelope phase, intensity
and duration of the pulse as control knobs. 
The additional freedom of quantum optimal control theory is important,
in particular when more complex systems are considered. 

The set of applications that we have presented here 
is far from being exhaustive, and our current
work opens many perspectives for both photoionization studies and
quantum optimal control theory. On the one hand, we have shown how to
develop optimization functionals that target directly an 
experimentally measurable quantity obtained from continuum
wavefunctions. On the other hand, since our approach is general, it
can straightforwardly be applied to  more complex examples. In this
respect it is desirable to lift the restriction to closed-shell
systems. This would pave the way to studying the role of electron
correlation in maximizing certain features in the photoelectron
spectrum.  Similarly, allowing for circular or elliptic polarization
of the electric field, one could envision, for example, to maximize
signatures of chirality in the photoelectron angular distributions.
This requires, however, substantial further development on the level
of the time-dependent electronic structure theory.

\begin{acknowledgments}
  Financial support by the State Hessen Initiative for the
  Development of Scientific and Economic Excellence (LOEWE) within the
  focus project Electron Dynamic of Chiral Systems (ELCH) 
  and the Deutsche Forschungsgemeinschaft
  under Grant No. SFB 925/A5 is gratefully acknowledged.
\end{acknowledgments}

\appendix
\section{Frequency and Amplitude Restriction}
\label{sec:frseqrestriction}

In the following, we present the derivation of Krotov's update equation for
the control, Eq.~\eqref{eq:mykrotova} using the approximation for 
$s(t)$ previously described. This allows for a more compact expression for
Krotov's equation for the specific constraints on the field used in 
this work. It is obtained following Ref.~\cite{ReichJCP12}: We seek to minimize 
the complete functional, Eq.~\eqref{eq:Jtotal}. 
In order to evaluate the extremum condition, we start by evaluating the 
functional derivative of the penalty functional with respect to the changes in
the control field $\mathbf{A}(t)$ in Eq.~\eqref{eq:Ca}, 
\begin{eqnarray}
  \dfrac{\delta C_a[\mathbf{A}]}{\delta\mathbf{A}(t)} &=& 2\lambda_a s^{-1}(t)
  \left(\mathbf{A}(t)- \mathbf{A}_{\rm ref}(t)\right).
\label{eq:tchikonovCa}
\end{eqnarray}
Next, we evaluate the functional derivative of Eq.~\eqref{eq:Comega}. Abbreviating 
$\omega^2\tilde{\gamma}(\omega)$
by $\tilde{h}(\omega)$ in  Eq.~\eqref{eq:Comega}, the functional derivative 
reads 
\begin{eqnarray} 
  \dfrac{\delta C_{\omega}[\mathbf{A}]}{\delta\mathbf{A}(t)} &=& 
 \lambda_{\omega} \int
 \mathbf{A}^{\star}(\omega)\dfrac{\delta\tilde{\mathbf{A}}(\omega)}{\delta
   \mathbf{A}(t)}\tilde{h}(\omega)\,d\omega \nonumber \\
   &&+\lambda_{\omega} \int
   \tilde{\mathbf{A}}(\omega)\dfrac{\delta\tilde{\mathbf{A}}^{\star}(\omega)}{\delta
     \mathbf{A}(t)}\tilde{h}(\omega)\,d\omega\,. 
\label{eq:omega0}
\end{eqnarray}
Using the fact that $\tilde{\mathbf{A}}(\omega)$ is the Fourier
transform of $\mathbf{A}(t)$,                       
\[
  \tilde{\mathbf{A}}(\omega)=
  \int \mathbf{A}(t) e^{-i\omega t}\,dt\,,
\]
the functional derivative becomes,             
\[ 
  \dfrac{\delta\tilde{\mathbf{A}}(\omega)}{\delta\mathbf{A}(t^\prime)} =
   e^{-i\omega t^\prime}\,, 
\]
such that 
\begin{eqnarray*}
\dfrac{\delta C_{\omega}[\mathbf{A}]}{\delta\mathbf{A}(t)} &=& 
\lambda_{\omega}\int \tilde{\mathbf{A}}^{\star}(\omega)e^{-i\omega t}\tilde{h}(\omega)\,d\omega\nn\\
 &&+ \lambda_{\omega}\int \tilde{\mathbf{A}}(\omega) e^{+i\omega t}\tilde{h}(\omega)\,d\omega\,\nn\,.
\end{eqnarray*}
This can be rewritten as 
\begin{eqnarray}
\label{eq:omega4}
\dfrac{\delta C_{\omega}[\mathbf{A}]}{\delta\mathbf{A}(t)} &=& 
 \lambda_{\omega}\int \tilde{\mathbf{A}}^{\star}(-\omega)e^{+i\omega
t}\tilde{h}(-\omega)\,d\omega\nn\\
&&+ \lambda_{\omega}\int \tilde{\mathbf{A}}(\omega) e^{+i\omega t}\tilde{h}(\omega)\,d\omega\,.
\end{eqnarray}
Since the control $\mathbf{A}(t)$ is a real function of time, 
$\tilde{\mathbf{A}}^{\star}(-\omega)=\tilde{\mathbf{A}}(\omega)$. 
Moreover, by construction
$\tilde{h}(\omega)=\tilde{h}(-\omega)$. Therefore, 
Eq.~\eqref{eq:omega4} becomes 
\begin{eqnarray}\label{eq:omega5}  
  \dfrac{\delta C_{\omega}[\mathbf{A}]}{\delta\mathbf{A}(t)} &=&
  2\lambda_{\omega}\int \tilde{\mathbf{A}}(\omega)e^{+i\omega t}\tilde{h}(\omega)\,d\omega \\
&=&2\lambda_{\omega}\int \tilde{\mathbf{A}}(\omega)e^{i\omega t}d\omega\int
h(\tau)e^{-i\omega\tau}\,d\tau\nonumber\\
& =&2\lambda_{\omega}\int h(\tau)d\tau\int \tilde{\mathbf{A}}(\omega)
e^{+i\omega(t-\tau)}d\omega \nonumber\\
&=& 2\tilde{\lambda}_{\omega}\int h(\tau)\mathbf{A}(t-\tau)d\tau = 2\tilde{\lambda}_{\omega}\mathbf{A}\star
h(t)\,, \nonumber
\end{eqnarray}
with $\tilde{\lambda}_{\omega} =\sqrt{2\pi}\lambda_{\omega}$, and $h(t)=\int{\tilde{h}(\omega)\exp{(+i\omega t})}\,d\omega/\sqrt{2\pi}$ 
and where $f\star g(t)$ refers to 
the convolution product of $f$ and $g$. 

We now calculate the functional derivative of the constraint
penalizing large values of $\dot{\mathbf{A}}(t)$, 
Eq.~\eqref{eq:Ce}. 
Assuming vanishing boundary conditions for $\mathbf{A}(t)$, we find, 
upon integration by parts,          
\begin{eqnarray}
  \dfrac{\delta C_e[\mathbf{A}]}{\delta\mathbf{A}(t)} &=& -2\lambda_e
  s^{-1}(t)\ddot{\mathbf{A}}(t)\,. 
\label{eq:tchikonov2}
\end{eqnarray}
Using Eqs.~\eqref{eq:tchikonovCa},~\eqref{eq:omega5} and~\eqref{eq:tchikonov2}, the 
extremum condition with respect to a variation in the 
control becomes  
\begin{eqnarray*}
0 & = & \lambda_a\,s^{-1}(t)\left(\mathbf{A}(t)-\mathbf{A}_{ref}(t)\right)  -\lambda_e\, s^{-1}(t)\ddot{\mathbf{A}}(t)   \nn\\
  & & + \tilde{\lambda}_{\omega}\, \mathbf{A}\star h(t) -
\mathfrak{Im}\left\{ \left\langle\chi(t)\left|\fr{\pa\hat{H}}{\mathbf{A}}
\right|\Psi(t)\right\rangle\right\}\nn\,.
\end{eqnarray*}
where the last term has been previously introduced in Eq.~\eqref{eq:mykrotov}. It can be
straightforwardly derived from variational principles, ie. Euler-Lagrange Lagrange
equation, or in the context of Pontriagin's maximum/minimum principle or in the
context of Krotov's optimization method, cf. Refs.~\cite{ReichJCP12,ReichJMO14}. It
stresses the dynamics to which the forward propagated state is subject to.
Solving for $\mathbf A(t)$ gives us the update rule for the optimized pulse,
\begin{eqnarray}
  \mathbf{A}(t) &= &\mathbf{A}_{ref}(t)
  +\fr{s(t)}{\lambda_a}\, 
  \mathfrak{Im}\left\{ \left\langle\chi(t)\left|\fr{\pa\hat{H}}{\mathbf{A}}
\right|\Psi(t)\right\rangle\right\} \nonumber\\ 
&&- \fr{\tilde{\lambda}_{\omega}}{\lambda_a} s(t)\mathbf{A}\star h(t)
+ \fr{\lambda_e}{\lambda}_a\ddot{\mathbf{A}}(t)\,, 
\label{eq:demo_mykrotov} 
\end{eqnarray}       
i.e., we retrieve Eq.~\eqref{eq:mykrotova}. Using the property 
\[
  \int \ddot{\mathbf{A}}(t)\, e^{-i\omega\,t}\, dt
  =-\omega^2\,\tilde{\mathbf{A}}(\omega)\,,\nn
\]
together with  Eq.~\eqref{choice_st} for $s(t)$, it is straighforward to write
Krotov's equation in frequency domain. To this end, we merely take the Fourier
transform of Eq.~\eqref{eq:demo_mykrotov} and utilize the well-known
property that the Fourier transform of a convolution of two functions
in time domain is the product of the functions in frequency domain. We
thus find 
\begin{eqnarray}\label{eq:demo_mykrotov_fourier}
  \tilde{\mathbf{A}}^{(k+1)}(\omega) &\approx&\tilde{\mathbf{A}}^{(k)}(\omega)
  + \tilde{I}^{(k+1)}(\omega) \\ \nonumber 
  &&- \fr{\tilde{\lambda}_{\omega}}{\lambda_a}\tilde{\mathbf{A}}^{(k+1)}(\omega) \tilde{h}(\omega)
- \omega^2\fr{\lambda_e}{\lambda_a}\tilde{\mathbf{A}}^{(k+1)}(\omega)  \,,
\end{eqnarray}
which yields Eq.~\eqref{eq:fourier}.


\begin{thebibliography}{49}
\expandafter\ifx\csname natexlab\endcsname\relax\def\natexlab#1{#1}\fi
\expandafter\ifx\csname bibnamefont\endcsname\relax
  \def\bibnamefont#1{#1}\fi
\expandafter\ifx\csname bibfnamefont\endcsname\relax
  \def\bibfnamefont#1{#1}\fi
\expandafter\ifx\csname citenamefont\endcsname\relax
  \def\citenamefont#1{#1}\fi
\expandafter\ifx\csname url\endcsname\relax
  \def\url#1{\texttt{#1}}\fi
\expandafter\ifx\csname urlprefix\endcsname\relax\def\urlprefix{URL }\fi
\providecommand{\bibinfo}[2]{#2}
\providecommand{\eprint}[2][]{\url{#2}}

\bibitem[{\citenamefont{H\"ufner}(2003)}]{Hufner}
\bibinfo{author}{\bibfnamefont{S.}~\bibnamefont{H\"ufner}},
  \emph{\bibinfo{title}{Photoelectron Spectroscopy, Principles and
  Applications}} (\bibinfo{publisher}{Springer}, \bibinfo{year}{2003}),
  \bibinfo{edition}{3rd} ed.

\bibitem[{\citenamefont{Meyer et~al.}(2010)\citenamefont{Meyer, Cubaynes,
  Richardson, Costello, Radcliffe, Li, D\"usterer, Fritzsche, Mihelic,
  Papamihail et~al.}}]{MeyerPRL10}
\bibinfo{author}{\bibfnamefont{M.}~\bibnamefont{Meyer}},
  \bibinfo{author}{\bibfnamefont{D.}~\bibnamefont{Cubaynes}},
  \bibinfo{author}{\bibfnamefont{V.}~\bibnamefont{Richardson}},
  \bibinfo{author}{\bibfnamefont{J.~T.} \bibnamefont{Costello}},
  \bibinfo{author}{\bibfnamefont{P.}~\bibnamefont{Radcliffe}},
  \bibinfo{author}{\bibfnamefont{W.~B.} \bibnamefont{Li}},
  \bibinfo{author}{\bibfnamefont{S.}~\bibnamefont{D\"usterer}},
  \bibinfo{author}{\bibfnamefont{S.}~\bibnamefont{Fritzsche}},
  \bibinfo{author}{\bibfnamefont{A.}~\bibnamefont{Mihelic}},
  \bibinfo{author}{\bibfnamefont{K.~G.} \bibnamefont{Papamihail}},
  \bibnamefont{et~al.}, \bibinfo{journal}{Phys. Rev. Lett.}
  \textbf{\bibinfo{volume}{104}}, \bibinfo{pages}{213001}
  (\bibinfo{year}{2010}).

\bibitem[{\citenamefont{Fabre et~al.}(1981)\citenamefont{Fabre, Agostini,
  Petite, and Clement}}]{FabreAtMolPhys81}
\bibinfo{author}{\bibfnamefont{F.}~\bibnamefont{Fabre}},
  \bibinfo{author}{\bibfnamefont{P.}~\bibnamefont{Agostini}},
  \bibinfo{author}{\bibfnamefont{G.}~\bibnamefont{Petite}}, \bibnamefont{and}
  \bibinfo{author}{\bibfnamefont{M.}~\bibnamefont{Clement}},
  \bibinfo{journal}{Journal of Physics B: Atomic and Molecular Physics}
  \textbf{\bibinfo{volume}{14}}, \bibinfo{pages}{L677} (\bibinfo{year}{1981}).

\bibitem[{\citenamefont{Becker and Shirley}(1996)}]{Becker}
\bibinfo{author}{\bibfnamefont{U.}~\bibnamefont{Becker}} \bibnamefont{and}
  \bibinfo{author}{\bibfnamefont{D.~A.} \bibnamefont{Shirley}},
  \emph{\bibinfo{title}{VUV and Soft X-Ray Photoionization}}
  (\bibinfo{publisher}{Springer Science \& Business Media},
  \bibinfo{year}{1996}), \bibinfo{edition}{3rd} ed.

\bibitem[{\citenamefont{Wu et~al.}(2011)\citenamefont{Wu, Hockett, and
  Stolow}}]{WuChemPhys11}
\bibinfo{author}{\bibfnamefont{G.}~\bibnamefont{Wu}},
  \bibinfo{author}{\bibfnamefont{P.}~\bibnamefont{Hockett}}, \bibnamefont{and}
  \bibinfo{author}{\bibfnamefont{A.}~\bibnamefont{Stolow}},
  \bibinfo{journal}{Phys. Chem. Chem. Phys.} \textbf{\bibinfo{volume}{13}},
  \bibinfo{pages}{18447} (\bibinfo{year}{2011}).

\bibitem[{\citenamefont{Blaga et~al.}(2009)\citenamefont{Blaga, Catoire,
  Colosimo, Paulus, Muller, Agostini, and DiMauro}}]{BlagaNat09}
\bibinfo{author}{\bibfnamefont{C.~I.} \bibnamefont{Blaga}},
  \bibinfo{author}{\bibfnamefont{F.}~\bibnamefont{Catoire}},
  \bibinfo{author}{\bibfnamefont{P.}~\bibnamefont{Colosimo}},
  \bibinfo{author}{\bibfnamefont{G.~G.} \bibnamefont{Paulus}},
  \bibinfo{author}{\bibfnamefont{H.~G.} \bibnamefont{Muller}},
  \bibinfo{author}{\bibfnamefont{P.}~\bibnamefont{Agostini}}, \bibnamefont{and}
  \bibinfo{author}{\bibfnamefont{L.}~\bibnamefont{DiMauro}},
  \bibinfo{journal}{Nature Phys.} \textbf{\bibinfo{volume}{5}},
  \bibinfo{pages}{335} (\bibinfo{year}{2009}).

\bibitem[{\citenamefont{Krausz and Ivanov}(2009)}]{KrauszModPhys09}
\bibinfo{author}{\bibfnamefont{F.}~\bibnamefont{Krausz}} \bibnamefont{and}
  \bibinfo{author}{\bibfnamefont{M.}~\bibnamefont{Ivanov}},
  \bibinfo{journal}{Rev. Mod. Phys.} \textbf{\bibinfo{volume}{81}},
  \bibinfo{pages}{163} (\bibinfo{year}{2009}).

\bibitem[{\citenamefont{Wabnitz et~al.}(2002)\citenamefont{Wabnitz, Bittner,
  de~Castro, Dohrmann, Gurtler, Laarmann, Laasch, Schulz, Swiderski, von
  Haeften et~al.}}]{WabnitzNAT02}
\bibinfo{author}{\bibfnamefont{H.}~\bibnamefont{Wabnitz}},
  \bibinfo{author}{\bibfnamefont{L.}~\bibnamefont{Bittner}},
  \bibinfo{author}{\bibfnamefont{A.~R.~B.} \bibnamefont{de~Castro}},
  \bibinfo{author}{\bibfnamefont{R.}~\bibnamefont{Dohrmann}},
  \bibinfo{author}{\bibfnamefont{P.}~\bibnamefont{Gurtler}},
  \bibinfo{author}{\bibfnamefont{T.}~\bibnamefont{Laarmann}},
  \bibinfo{author}{\bibfnamefont{W.}~\bibnamefont{Laasch}},
  \bibinfo{author}{\bibfnamefont{J.}~\bibnamefont{Schulz}},
  \bibinfo{author}{\bibfnamefont{A.}~\bibnamefont{Swiderski}},
  \bibinfo{author}{\bibfnamefont{K.}~\bibnamefont{von Haeften}},
  \bibnamefont{et~al.}, \bibinfo{journal}{Nature Physics}
  \textbf{\bibinfo{volume}{420}}, \bibinfo{pages}{482} (\bibinfo{year}{2002}).

\bibitem[{\citenamefont{Corkum and Krausz}(2007)}]{CorkumNAT07}
\bibinfo{author}{\bibfnamefont{P.}~\bibnamefont{Corkum}} \bibnamefont{and}
  \bibinfo{author}{\bibfnamefont{F.}~\bibnamefont{Krausz}},
  \bibinfo{journal}{Nature Physics} \textbf{\bibinfo{volume}{3}},
  \bibinfo{pages}{381} (\bibinfo{year}{2007}).

\bibitem[{\citenamefont{Kanter et~al.}(2011)\citenamefont{Kanter, Kr\"assig,
  Li, March, Ho, Rohringer, Santra, Southworth, DiMauro, Doumy
  et~al.}}]{KanterPRL11}
\bibinfo{author}{\bibfnamefont{E.~P.} \bibnamefont{Kanter}},
  \bibinfo{author}{\bibfnamefont{B.}~\bibnamefont{Kr\"assig}},
  \bibinfo{author}{\bibfnamefont{Y.}~\bibnamefont{Li}},
  \bibinfo{author}{\bibfnamefont{A.~M.} \bibnamefont{March}},
  \bibinfo{author}{\bibfnamefont{P.}~\bibnamefont{Ho}},
  \bibinfo{author}{\bibfnamefont{N.}~\bibnamefont{Rohringer}},
  \bibinfo{author}{\bibfnamefont{R.}~\bibnamefont{Santra}},
  \bibinfo{author}{\bibfnamefont{S.~H.} \bibnamefont{Southworth}},
  \bibinfo{author}{\bibfnamefont{L.~F.} \bibnamefont{DiMauro}},
  \bibinfo{author}{\bibfnamefont{G.}~\bibnamefont{Doumy}},
  \bibnamefont{et~al.}, \bibinfo{journal}{Phys. Rev. Lett.}
  \textbf{\bibinfo{volume}{107}}, \bibinfo{pages}{233001}
  (\bibinfo{year}{2011}).

\bibitem[{\citenamefont{Schwarz}(1980)}]{SchwarzElecSpectros80}
\bibinfo{author}{\bibfnamefont{H.~E.} \bibnamefont{Schwarz}},
  \bibinfo{journal}{Journal of Electron Spectroscopy and Related Phenomena}
  \textbf{\bibinfo{volume}{21}} (\bibinfo{year}{1980}).

\bibitem[{\citenamefont{Duinker}(1982)}]{DuinkerRMP82}
\bibinfo{author}{\bibfnamefont{P.}~\bibnamefont{Duinker}},
  \bibinfo{journal}{Rev. Mod. Phys.} \textbf{\bibinfo{volume}{54}},
  \bibinfo{pages}{325} (\bibinfo{year}{1982}).

\bibitem[{\citenamefont{Reid}(2003)}]{ReidARPC03}
\bibinfo{author}{\bibfnamefont{K.~L.} \bibnamefont{Reid}},
  \bibinfo{journal}{Annu. Rev. Phys. Chem.} \textbf{\bibinfo{volume}{54}},
  \bibinfo{pages}{397} (\bibinfo{year}{2003}).

\bibitem[{\citenamefont{Meier and Engel}(1994)}]{MeierPRL94}
\bibinfo{author}{\bibfnamefont{C.}~\bibnamefont{Meier}} \bibnamefont{and}
  \bibinfo{author}{\bibfnamefont{V.}~\bibnamefont{Engel}},
  \bibinfo{journal}{Phys. Rev. Lett.} \textbf{\bibinfo{volume}{73}},
  \bibinfo{pages}{3207} (\bibinfo{year}{1994}).

\bibitem[{\citenamefont{Shen and Engel}(2002)}]{ShenCPL02}
\bibinfo{author}{\bibfnamefont{Z.}~\bibnamefont{Shen}} \bibnamefont{and}
  \bibinfo{author}{\bibfnamefont{V.}~\bibnamefont{Engel}},
  \bibinfo{journal}{Chem. Phys. Lett.} \textbf{\bibinfo{volume}{358}},
  \bibinfo{pages}{344 } (\bibinfo{year}{2002}).

\bibitem[{\citenamefont{Wollenhaupt et~al.}(2005)\citenamefont{Wollenhaupt,
  Engel, and Baumert}}]{WollenhauptARPC05}
\bibinfo{author}{\bibfnamefont{M.}~\bibnamefont{Wollenhaupt}},
  \bibinfo{author}{\bibfnamefont{V.}~\bibnamefont{Engel}}, \bibnamefont{and}
  \bibinfo{author}{\bibfnamefont{T.}~\bibnamefont{Baumert}},
  \bibinfo{journal}{Annu. Rev. Phys. Chem.} \textbf{\bibinfo{volume}{56}},
  \bibinfo{pages}{25} (\bibinfo{year}{2005}).

\bibitem[{\citenamefont{Gr\"afe et~al.}(2005)\citenamefont{Gr\"afe, Erdmann,
  and Engel}}]{GraefePRA05}
\bibinfo{author}{\bibfnamefont{S.}~\bibnamefont{Gr\"afe}},
  \bibinfo{author}{\bibfnamefont{M.}~\bibnamefont{Erdmann}}, \bibnamefont{and}
  \bibinfo{author}{\bibfnamefont{V.}~\bibnamefont{Engel}},
  \bibinfo{journal}{Phys. Rev. A} \textbf{\bibinfo{volume}{72}},
  \bibinfo{pages}{013404} (\bibinfo{year}{2005}).

\bibitem[{\citenamefont{Braun et~al.}(2014)\citenamefont{Braun, Bayer, Sarpe,
  Siemering, de~Vivie-Riedle, Baumert, and Wollenhaupt}}]{BraunJPB14}
\bibinfo{author}{\bibfnamefont{H.}~\bibnamefont{Braun}},
  \bibinfo{author}{\bibfnamefont{T.}~\bibnamefont{Bayer}},
  \bibinfo{author}{\bibfnamefont{C.}~\bibnamefont{Sarpe}},
  \bibinfo{author}{\bibfnamefont{R.}~\bibnamefont{Siemering}},
  \bibinfo{author}{\bibfnamefont{R.}~\bibnamefont{de~Vivie-Riedle}},
  \bibinfo{author}{\bibfnamefont{T.}~\bibnamefont{Baumert}}, \bibnamefont{and}
  \bibinfo{author}{\bibfnamefont{M.}~\bibnamefont{Wollenhaupt}},
  \bibinfo{journal}{J. Phys. B} \textbf{\bibinfo{volume}{47}},
  \bibinfo{pages}{124015} (\bibinfo{year}{2014}).

\bibitem[{\citenamefont{Reich et~al.}(2012)\citenamefont{Reich, Ndong, and
  Koch}}]{ReichJCP12}
\bibinfo{author}{\bibfnamefont{D.~M.} \bibnamefont{Reich}},
  \bibinfo{author}{\bibfnamefont{M.}~\bibnamefont{Ndong}}, \bibnamefont{and}
  \bibinfo{author}{\bibfnamefont{C.~P.} \bibnamefont{Koch}},
  \bibinfo{journal}{J. Chem. Phys.} \textbf{\bibinfo{volume}{136}},
  \bibinfo{pages}{104103} (\bibinfo{year}{2012}).

\bibitem[{\citenamefont{Greenman et~al.}(2010)\citenamefont{Greenman, Ho,
  Pabst, Kamarchik, Mazziotti, and Santra}}]{GreenmanPRA10}
\bibinfo{author}{\bibfnamefont{L.}~\bibnamefont{Greenman}},
  \bibinfo{author}{\bibfnamefont{P.~J.} \bibnamefont{Ho}},
  \bibinfo{author}{\bibfnamefont{S.}~\bibnamefont{Pabst}},
  \bibinfo{author}{\bibfnamefont{E.}~\bibnamefont{Kamarchik}},
  \bibinfo{author}{\bibfnamefont{D.~A.} \bibnamefont{Mazziotti}},
  \bibnamefont{and} \bibinfo{author}{\bibfnamefont{R.}~\bibnamefont{Santra}},
  \bibinfo{journal}{Phys. Rev. A} \textbf{\bibinfo{volume}{82}},
  \bibinfo{pages}{023406} (\bibinfo{year}{2010}).

\bibitem[{\citenamefont{Karamatskou et~al.}(2014)\citenamefont{Karamatskou,
  Pabst, Chen, and Santra}}]{AntoniaPRA14}
\bibinfo{author}{\bibfnamefont{A.}~\bibnamefont{Karamatskou}},
  \bibinfo{author}{\bibfnamefont{S.}~\bibnamefont{Pabst}},
  \bibinfo{author}{\bibfnamefont{Y.-J.} \bibnamefont{Chen}}, \bibnamefont{and}
  \bibinfo{author}{\bibfnamefont{R.}~\bibnamefont{Santra}},
  \bibinfo{journal}{Phys. Rev. A} \textbf{\bibinfo{volume}{89}},
  \bibinfo{pages}{033415} (\bibinfo{year}{2014}).

\bibitem[{\citenamefont{Karamatskou et~al.}(2015)\citenamefont{Karamatskou,
  Pabst, Chen, and Santra}}]{AntoniaPRA15Erratum}
\bibinfo{author}{\bibfnamefont{A.}~\bibnamefont{Karamatskou}},
  \bibinfo{author}{\bibfnamefont{S.}~\bibnamefont{Pabst}},
  \bibinfo{author}{\bibfnamefont{Y.-J.} \bibnamefont{Chen}}, \bibnamefont{and}
  \bibinfo{author}{\bibfnamefont{R.}~\bibnamefont{Santra}},
  \bibinfo{journal}{Phys. Rev. A} \textbf{\bibinfo{volume}{91}},
  \bibinfo{pages}{069907} (\bibinfo{year}{2015}).

\bibitem[{\citenamefont{Klamroth}(2006)}]{KlamrothJCP06}
\bibinfo{author}{\bibfnamefont{T.}~\bibnamefont{Klamroth}},
  \bibinfo{journal}{J. Chem. Phys.} \textbf{\bibinfo{volume}{124}},
  \bibinfo{eid}{144310} (\bibinfo{year}{2006}).

\bibitem[{\citenamefont{Mundt and Tannor}(2009)}]{MundtNJP09}
\bibinfo{author}{\bibfnamefont{M.}~\bibnamefont{Mundt}} \bibnamefont{and}
  \bibinfo{author}{\bibfnamefont{D.~J.} \bibnamefont{Tannor}},
  \bibinfo{journal}{New J. Phys.} \textbf{\bibinfo{volume}{11}},
  \bibinfo{pages}{105038} (\bibinfo{year}{2009}).

\bibitem[{\citenamefont{Castro et~al.}(2012)\citenamefont{Castro, Werschnik,
  and Gross}}]{CastroPRL12}
\bibinfo{author}{\bibfnamefont{A.}~\bibnamefont{Castro}},
  \bibinfo{author}{\bibfnamefont{J.}~\bibnamefont{Werschnik}},
  \bibnamefont{and} \bibinfo{author}{\bibfnamefont{E.~K.~U.}
  \bibnamefont{Gross}}, \bibinfo{journal}{Phys. Rev. Lett.}
  \textbf{\bibinfo{volume}{109}}, \bibinfo{pages}{153603}
  (\bibinfo{year}{2012}).

\bibitem[{\citenamefont{Hellgren et~al.}(2013)\citenamefont{Hellgren,
  R\"as\"anen, and Gross}}]{HellgrenPRA13}
\bibinfo{author}{\bibfnamefont{M.}~\bibnamefont{Hellgren}},
  \bibinfo{author}{\bibfnamefont{E.}~\bibnamefont{R\"as\"anen}},
  \bibnamefont{and} \bibinfo{author}{\bibfnamefont{E.~K.~U.}
  \bibnamefont{Gross}}, \bibinfo{journal}{Phys. Rev. A}
  \textbf{\bibinfo{volume}{88}}, \bibinfo{pages}{013414}
  (\bibinfo{year}{2013}).

\bibitem[{\citenamefont{McCurdy and Rescigno}(1997)}]{McCurdyPRA97}
\bibinfo{author}{\bibfnamefont{C.~W.} \bibnamefont{McCurdy}} \bibnamefont{and}
  \bibinfo{author}{\bibfnamefont{T.~N.} \bibnamefont{Rescigno}},
  \bibinfo{journal}{Phys. Rev. A} \textbf{\bibinfo{volume}{56}},
  \bibinfo{pages}{R4369} (\bibinfo{year}{1997}).

\bibitem[{\citenamefont{Mart\'in}(1999)}]{MartinJPhysB99}
\bibinfo{author}{\bibfnamefont{F.}~\bibnamefont{Mart\'in}},
  \bibinfo{journal}{Journal of Physics B: Atomic, Molecular and Optical
  Physics} \textbf{\bibinfo{volume}{32}}, \bibinfo{pages}{R197}
  (\bibinfo{year}{1999}).

\bibitem[{\citenamefont{Rescigno and McCurdy}(2000)}]{RescignoPRA2000}
\bibinfo{author}{\bibfnamefont{T.~N.} \bibnamefont{Rescigno}} \bibnamefont{and}
  \bibinfo{author}{\bibfnamefont{C.~W.} \bibnamefont{McCurdy}},
  \bibinfo{journal}{Phys. Rev. A} \textbf{\bibinfo{volume}{62}},
  \bibinfo{pages}{032706} (\bibinfo{year}{2000}).

\bibitem[{\citenamefont{Bachau et~al.}(2001)\citenamefont{Bachau, Cormier,
  Decleva, Hansen, and Mart\'in}}]{bachau01}
\bibinfo{author}{\bibfnamefont{H.}~\bibnamefont{Bachau}},
  \bibinfo{author}{\bibfnamefont{E.}~\bibnamefont{Cormier}},
  \bibinfo{author}{\bibfnamefont{P.}~\bibnamefont{Decleva}},
  \bibinfo{author}{\bibfnamefont{J.~E.} \bibnamefont{Hansen}},
  \bibnamefont{and} \bibinfo{author}{\bibfnamefont{F.}~\bibnamefont{Mart\'in}},
  \bibinfo{journal}{Reports on Progress in Physics}
  \textbf{\bibinfo{volume}{64}}, \bibinfo{pages}{1815} (\bibinfo{year}{2001}).

\bibitem[{\citenamefont{McCurdy and Mart\'in}(2004)}]{McCurdyJPhysB04}
\bibinfo{author}{\bibfnamefont{C.~W.} \bibnamefont{McCurdy}} \bibnamefont{and}
  \bibinfo{author}{\bibfnamefont{F.}~\bibnamefont{Mart\'in}},
  \bibinfo{journal}{Journal of Physics B: Atomic, Molecular and Optical
  Physics} \textbf{\bibinfo{volume}{37}}, \bibinfo{pages}{917}
  (\bibinfo{year}{2004}).

\bibitem[{\citenamefont{Greenman et~al.}(2014)\citenamefont{Greenman, Koch, and
  Whaley}}]{Greenman}
\bibinfo{author}{\bibfnamefont{L.}~\bibnamefont{Greenman}},
  \bibinfo{author}{\bibfnamefont{C.~P.} \bibnamefont{Koch}}, \bibnamefont{and}
  \bibinfo{author}{\bibfnamefont{K.~B.} \bibnamefont{Whaley}},
  \bibinfo{journal}{Phys. Rev. A} \textbf{\bibinfo{volume}{92}},
  \bibinfo{pages}{013407} (\bibinfo{year}{2015}).

\bibitem[{\citenamefont{Rathje et~al.}(2013)\citenamefont{Rathje, Sayler, Zeng,
  Wustelt, Figger, Esry, and Paulus}}]{RathjePRL13}
\bibinfo{author}{\bibfnamefont{T.}~\bibnamefont{Rathje}},
  \bibinfo{author}{\bibfnamefont{A.~M.} \bibnamefont{Sayler}},
  \bibinfo{author}{\bibfnamefont{S.}~\bibnamefont{Zeng}},
  \bibinfo{author}{\bibfnamefont{P.}~\bibnamefont{Wustelt}},
  \bibinfo{author}{\bibfnamefont{H.}~\bibnamefont{Figger}},
  \bibinfo{author}{\bibfnamefont{B.~D.} \bibnamefont{Esry}}, \bibnamefont{and}
  \bibinfo{author}{\bibfnamefont{G.~G.} \bibnamefont{Paulus}},
  \bibinfo{journal}{Phys. Rev. Lett.} \textbf{\bibinfo{volume}{111}},
  \bibinfo{pages}{093002} (\bibinfo{year}{2013}).

\bibitem[{\citenamefont{Shvetsov-Shilovski
  et~al.}(2014)\citenamefont{Shvetsov-Shilovski, R\"as\"anen, Paulus, and
  Madsen}}]{ShvetsovPRA14}
\bibinfo{author}{\bibfnamefont{N.~I.} \bibnamefont{Shvetsov-Shilovski}},
  \bibinfo{author}{\bibfnamefont{E.}~\bibnamefont{R\"as\"anen}},
  \bibinfo{author}{\bibfnamefont{G.~G.} \bibnamefont{Paulus}},
  \bibnamefont{and} \bibinfo{author}{\bibfnamefont{L.~B.}
  \bibnamefont{Madsen}}, \bibinfo{journal}{Phys. Rev. A}
  \textbf{\bibinfo{volume}{89}} (\bibinfo{year}{2014}).

\bibitem[{\citenamefont{Pabst et~al.}(Rev 1425, 2014)\citenamefont{Pabst,
  Greenman, and Santra}}]{xcid}
\bibinfo{author}{\bibfnamefont{S.}~\bibnamefont{Pabst}},
  \bibinfo{author}{\bibfnamefont{L.}~\bibnamefont{Greenman}}, \bibnamefont{and}
  \bibinfo{author}{\bibfnamefont{R.}~\bibnamefont{Santra}},
  \emph{\bibinfo{title}{{\textsc{XCID}} program package for multichannel
  ionization dynamics}} (\bibinfo{year}{Rev 1425, 2014}).

\bibitem[{\citenamefont{Leforestier et~al.}(1991)\citenamefont{Leforestier,
  Bisseling, Cerjan, Feit, Friesner, Guldberg, Hammerich, Jolicard, Karrlein,
  Meyer et~al.}}]{leforestierJCP91}
\bibinfo{author}{\bibfnamefont{C.}~\bibnamefont{Leforestier}},
  \bibinfo{author}{\bibfnamefont{R.}~\bibnamefont{Bisseling}},
  \bibinfo{author}{\bibfnamefont{C.}~\bibnamefont{Cerjan}},
  \bibinfo{author}{\bibfnamefont{M.}~\bibnamefont{Feit}},
  \bibinfo{author}{\bibfnamefont{R.}~\bibnamefont{Friesner}},
  \bibinfo{author}{\bibfnamefont{A.}~\bibnamefont{Guldberg}},
  \bibinfo{author}{\bibfnamefont{A.}~\bibnamefont{Hammerich}},
  \bibinfo{author}{\bibfnamefont{G.}~\bibnamefont{Jolicard}},
  \bibinfo{author}{\bibfnamefont{W.}~\bibnamefont{Karrlein}},
  \bibinfo{author}{\bibfnamefont{H.-D.} \bibnamefont{Meyer}},
  \bibnamefont{et~al.}, \bibinfo{journal}{J. Comput. Phys.}
  \textbf{\bibinfo{volume}{94}}, \bibinfo{pages}{59 } (\bibinfo{year}{1991}).

\bibitem[{\citenamefont{Kosloff}(1994)}]{kosloffRPC94}
\bibinfo{author}{\bibfnamefont{R.}~\bibnamefont{Kosloff}},
  \bibinfo{journal}{Annu. Rev. Phys. Chem.} \textbf{\bibinfo{volume}{45}},
  \bibinfo{pages}{145} (\bibinfo{year}{1994}).

\bibitem[{\citenamefont{Rohringer et~al.}(2006)\citenamefont{Rohringer, Gordon,
  and Santra}}]{RohPRA06}
\bibinfo{author}{\bibfnamefont{N.}~\bibnamefont{Rohringer}},
  \bibinfo{author}{\bibfnamefont{A.}~\bibnamefont{Gordon}}, \bibnamefont{and}
  \bibinfo{author}{\bibfnamefont{R.}~\bibnamefont{Santra}},
  \bibinfo{journal}{Phys. Rev. A} \textbf{\bibinfo{volume}{74}},
  \bibinfo{pages}{043420} (\bibinfo{year}{2006}).

\bibitem[{\citenamefont{Tong et~al.}(2006)\citenamefont{Tong, Hino, and
  Toshima}}]{TongPRA06}
\bibinfo{author}{\bibfnamefont{X.~M.} \bibnamefont{Tong}},
  \bibinfo{author}{\bibfnamefont{K.}~\bibnamefont{Hino}}, \bibnamefont{and}
  \bibinfo{author}{\bibfnamefont{N.}~\bibnamefont{Toshima}},
  \bibinfo{journal}{Phys. Rev. A} \textbf{\bibinfo{volume}{74}},
  \bibinfo{pages}{031405} (\bibinfo{year}{2006}).

\bibitem[{\citenamefont{Palao et~al.}(2013)\citenamefont{Palao, Reich, and
  Koch}}]{PalaoPRA13}
\bibinfo{author}{\bibfnamefont{J.~P.} \bibnamefont{Palao}},
  \bibinfo{author}{\bibfnamefont{D.~M.} \bibnamefont{Reich}}, \bibnamefont{and}
  \bibinfo{author}{\bibfnamefont{C.~P.} \bibnamefont{Koch}},
  \bibinfo{journal}{Phys. Rev. A} \textbf{\bibinfo{volume}{88}},
  \bibinfo{pages}{053409} (\bibinfo{year}{2013}).

\bibitem[{\citenamefont{Reich et~al.}(2014)\citenamefont{Reich, Palao, and
  Koch}}]{ReichJMO14}
\bibinfo{author}{\bibfnamefont{D.~M.} \bibnamefont{Reich}},
  \bibinfo{author}{\bibfnamefont{J.~P.} \bibnamefont{Palao}}, \bibnamefont{and}
  \bibinfo{author}{\bibfnamefont{C.~P.} \bibnamefont{Koch}},
  \bibinfo{journal}{J. Mod. Opt.} \textbf{\bibinfo{volume}{61}},
  \bibinfo{pages}{822} (\bibinfo{year}{2014}).

\bibitem[{\citenamefont{Hohenester et~al.}(2007)\citenamefont{Hohenester,
  Rekdal, Borz\`\i, and Schmiedmayer}}]{HohenesterPRA07}
\bibinfo{author}{\bibfnamefont{U.}~\bibnamefont{Hohenester}},
  \bibinfo{author}{\bibfnamefont{P.~K.} \bibnamefont{Rekdal}},
  \bibinfo{author}{\bibfnamefont{A.}~\bibnamefont{Borz\`\i}}, \bibnamefont{and}
  \bibinfo{author}{\bibfnamefont{J.}~\bibnamefont{Schmiedmayer}},
  \bibinfo{journal}{Phys. Rev. A} \textbf{\bibinfo{volume}{75}},
  \bibinfo{pages}{023602} (\bibinfo{year}{2007}).

\bibitem[{\citenamefont{Palao and Kosloff}(2003)}]{PalaoPRA03}
\bibinfo{author}{\bibfnamefont{J.~P.} \bibnamefont{Palao}} \bibnamefont{and}
  \bibinfo{author}{\bibfnamefont{R.}~\bibnamefont{Kosloff}},
  \bibinfo{journal}{Phys. Rev. A} \textbf{\bibinfo{volume}{68}},
  \bibinfo{pages}{062308} (\bibinfo{year}{2003}).

\bibitem[{\citenamefont{Eitan et~al.}(2011)\citenamefont{Eitan, Mundt, and
  Tannor}}]{EitanPRA11}
\bibinfo{author}{\bibfnamefont{R.}~\bibnamefont{Eitan}},
  \bibinfo{author}{\bibfnamefont{M.}~\bibnamefont{Mundt}}, \bibnamefont{and}
  \bibinfo{author}{\bibfnamefont{D.~J.} \bibnamefont{Tannor}},
  \bibinfo{journal}{Phys. Rev. A} \textbf{\bibinfo{volume}{83}},
  \bibinfo{pages}{053426} (\bibinfo{year}{2011}).

\bibitem[{\citenamefont{J\"ager et~al.}(2014)\citenamefont{J\"ager, Reich,
  Goerz, Koch, and Hohenester}}]{JaegerPRA14}
\bibinfo{author}{\bibfnamefont{G.}~\bibnamefont{J\"ager}},
  \bibinfo{author}{\bibfnamefont{D.~M.} \bibnamefont{Reich}},
  \bibinfo{author}{\bibfnamefont{M.~H.} \bibnamefont{Goerz}},
  \bibinfo{author}{\bibfnamefont{C.~P.} \bibnamefont{Koch}}, \bibnamefont{and}
  \bibinfo{author}{\bibfnamefont{U.}~\bibnamefont{Hohenester}},
  \bibinfo{journal}{Phys. Rev. A} \textbf{\bibinfo{volume}{90}},
  \bibinfo{pages}{033628} (\bibinfo{year}{2014}).

\bibitem[{\citenamefont{Schirmer and de~Fouquieres}(2009)}]{SophieNJP2011}
\bibinfo{author}{\bibfnamefont{S.~G.} \bibnamefont{Schirmer}} \bibnamefont{and}
  \bibinfo{author}{\bibfnamefont{P.}~\bibnamefont{de~Fouquieres}},
  \bibinfo{journal}{New J. Phys.} \textbf{\bibinfo{volume}{13}},
  \bibinfo{pages}{073029 (35pp)} (\bibinfo{year}{2009}).

\bibitem[{\citenamefont{Schr\"oder and Brown}(2009)}]{SchroederNJP09}
\bibinfo{author}{\bibfnamefont{M.}~\bibnamefont{Schr\"oder}} \bibnamefont{and}
  \bibinfo{author}{\bibfnamefont{A.}~\bibnamefont{Brown}},
  \bibinfo{journal}{New J. Phys.} \textbf{\bibinfo{volume}{11}},
  \bibinfo{pages}{105031 (13pp)} (\bibinfo{year}{2009}).

\bibitem[{\citenamefont{Lapert et~al.}(2009)\citenamefont{Lapert, Tehini,
  Turinici, and Sugny}}]{LapertPRA09}
\bibinfo{author}{\bibfnamefont{M.}~\bibnamefont{Lapert}},
  \bibinfo{author}{\bibfnamefont{R.}~\bibnamefont{Tehini}},
  \bibinfo{author}{\bibfnamefont{G.}~\bibnamefont{Turinici}}, \bibnamefont{and}
  \bibinfo{author}{\bibfnamefont{D.}~\bibnamefont{Sugny}},
  \bibinfo{journal}{Phys. Rev. A} \textbf{\bibinfo{volume}{79}},
  \bibinfo{pages}{063411} (\bibinfo{year}{2009}).

\bibitem[{\citenamefont{Chelkowski et~al.}(2004)\citenamefont{Chelkowski,
  Bandrauk, and Apolonski}}]{ChelkowskiPRA2004}
\bibinfo{author}{\bibfnamefont{S.}~\bibnamefont{Chelkowski}},
  \bibinfo{author}{\bibfnamefont{A.~D.} \bibnamefont{Bandrauk}},
  \bibnamefont{and}
  \bibinfo{author}{\bibfnamefont{A.}~\bibnamefont{Apolonski}},
  \bibinfo{journal}{Phys. Rev. A} \textbf{\bibinfo{volume}{70}},
  \bibinfo{pages}{013815} (\bibinfo{year}{2004}).

\end{thebibliography}

\end{document}